\documentclass[epsfig,12pt]{article}
\usepackage{amsfonts}
\usepackage{amssymb}
\usepackage{graphicx}
\usepackage{amsmath}

\input epsf.sty
\newtheorem{theorem}{Theorem}
\newtheorem{acknowledgement}[theorem]{Acknowledgement}

\input{tcilatex}
\makeatletter \@addtoreset{equation}{section}
\renewcommand{\theequation}{\thesection.\arabic{equation}}
\begin{document}

\title{%
\rightline{\mbox{\normalsize
{Lab/UFR-HEP0605/GNPHE/0605/VACBT/0605}}} \textbf{Pure fermionic\ twistor
like model }\\
\textbf{\& target space supersymmetry}}
\author{R. Ahl Laamara, L.B Drissi, H. Jehjouh, E H Saidi\thanks{%
h-saidi@fsr.ac.ma} \\
{\small \textit{1.Lab/UFR- Physique des Hautes Energies, Facul\'{e} des
Sciences, Rabat, Morocco,}}\\
{\small \textit{2.GNPHE, Groupement National de Physique des Hautes
Energies, }}\\
{\small \textit{Si\`{e}ge focal: Facult\'{e} des Sciences de Rabat, Morocco.}%
}\\
{\small \textit{3.Virtual African Centre for Basic Science and Technology,
VACBT, }}\\
{\small \textit{Focal point, LabUFR-PHE, Rabat, Morocco.}}}
\maketitle

\begin{abstract}
Using world line fermions $\Upsilon _{\pm }^{m}=\Upsilon _{\pm }^{m}\left(
\tau \right) $ valued in vector representation of $SO\left( d,4-d\right) $
with $d=2,3,4,$ we develop a pure fermionic analog of Penrose twistor
construction. First, we show that Fermi antisymmetry requiring $\left(
\Upsilon _{\pm }^{m}\right) ^{2}=0$ can be solved by using twistor like
variables. Then we study the corresponding dual twistor like field action
and show that quantum spectrum exhibits naturally $4D$ $\mathcal{N}=1$
target space supersymmetry. Higher spin world line field solutions of the
constraint $\left( \Pi _{s}^{m}\right) ^{2}=0$, $s\in \mathbb{Z}$ are also
discussed.\newline
\bigskip

\end{abstract}


\newpage

\section{Introduction}

\qquad During the few last years, a lot of developments has been made on
twistor approach for describing $4D$ $\mathcal{N}=4$ supersymmetric
Yang-Mills and $10D$ superstring theories in complex projective superspace $%
\mathbb{CP}^{(3|4)}$ $\cite{1}$-$\cite{5}$, see also \textrm{\cite{6}-\cite%
{10}} for earlier related works. An important interest has been given to the
study of twistor supersymmetric gauge theory and stringy extension $\cite{11}
$-$\cite{15}$. The key issue behind this revival of interest in Penrose
model \textrm{\cite{16}-\cite{18}} is the proposal of $\cite{1}$ where it
was shown that the supertwistor space $\mathbb{CP}^{(3|4)}$ is a Calabi-Yau
supermanifold $\cite{19}$-$\cite{28}$ capturing naturally special properties
of gauge theory correlation functions; in particular planar amplitudes of $4D
$ $\mathcal{N}=4$ supersymmetric U($N$) Yang-Mills theory, and exhibiting
the good properties of a target space for topological string theory on
Calabi-Yau threefolds \textrm{\cite{29}-} $\cite{35}$.

\qquad Motivated by recent results in this matter and particular aspects of
the so called two-time physics $\cite{36}$, we propose in this paper to set
up the basis of a pure world line fermionic extension of standard Penrose
model. Our construction is based on solving the nilpotency property $\left(
\Upsilon _{m}\right) ^{2}=0$ of the world line fermions $\Upsilon
_{m}=\Upsilon _{m}\left( \tau \right) $ by using a Penrose like method.
Recall that twistor approach of Penrose is based, amongst others, on solving
the condition $\left( p_{m}\right) ^{2}=0$ of a massless particle in 4D-
space time. It happens that, besides the striking analogy between the
solutions of $\left( p_{m}\right) ^{2}=0$ and $\left( \Upsilon _{m}\right)
^{2}=0$ and their underlying twistor world line field formulations, we find
moreover that fermionic twistor like space is a graded superspace capturing
four dimension target space supersymmetry in a wonderful way.

\qquad To derive our pure fermionic twistor like model, we shall mimick the
standard twistor derivation of Penrose $\cite{16}$ as reformulated in $\cite%
{1}$. The main steps of this construction can, roughly, be summarized as
follows:\newline
(\textbf{a}) The usual real world line free massless field $x_{\alpha }^{%
\dot{\alpha}}\left( \tau \right) $ of standard Penrose twistor model is now
replaced by a free world line fermion $\Upsilon _{\alpha }^{\dot{\alpha}%
}\left( \tau \right) $. Both of these world line fields are 4-vectors of the
real $\mathbb{R}^{\left( \mathrm{d,4-d}\right) }$ space, $d=2,3,4$, with
isometry group $SO\left( d,4-d\right) $ and diagonal metric $\eta _{mn}$
with $\left( d,4-d\right) $ signature.\newline
(\textbf{b}) The role of the conjugate momentum $p_{\dot{\alpha}}^{\alpha
}=\left( \frac{\delta S_{BT}}{\delta \partial _{\tau }x_{\alpha }^{\dot{%
\alpha}}}\right) $ of the field $x_{\alpha }^{\dot{\alpha}}$ gets played by $%
{\Large \pi }_{\dot{\alpha}}^{\alpha }=\left( \frac{\delta S_{FT}}{\delta
\partial _{\tau }\Upsilon _{\alpha }^{\dot{\alpha}}}\right) $, the fermionic
conjugate momentum of $\Upsilon _{\alpha }^{\dot{\alpha}}$. In these
relations, $\mathcal{S}_{BT}=\mathcal{S}\left[ X_{\alpha }^{\dot{\alpha}},p_{%
\dot{\alpha}}^{\alpha }\right] $ and $\mathcal{S}_{FT}=\mathcal{S}\left[
\Upsilon _{\alpha }^{\dot{\alpha}},{\Large \pi }_{\dot{\alpha}}^{\alpha }%
\right] $ stand for the world line field actions of the bosonic $x_{\alpha
}^{\dot{\alpha}}$, $p_{\dot{\alpha}}^{\alpha }$ and the fermionic $\Upsilon
_{\alpha }^{\dot{\alpha}},$ ${\Large \pi }_{\dot{\alpha}}^{\alpha }$ fields
respectively.\newline
(\textbf{c}) The role of the zero mass relation $p^{2}=0$ of the free field
bosonic theory is now played by the fermionic nilpotency property ${\Large %
\pi }^{2}=0$ of the fermionic field ${\Large \pi }_{\dot{\alpha}}^{\alpha }$%
. Notice that in our fermionic analog, there is a subtlety to announce.
Besides \textrm{Tr}$\left( {\Large \pi }^{2}\right) =0$, there are two more
constraint eqs one has to deal with. They are given by \textrm{Tr}$\left(
\Upsilon ^{2}\right) =0$ and \textrm{Tr}$\left( {\Large \pi }\Upsilon
\right) =0$ and it turns out that they are altogether captured by a local $%
SP\left( 2,\mathbb{R}\right) $ gauge invariance to be discussed in section 4.

The next step is to solve the constraint eqs using Penrose method and derive
the dual twistor field action describing the fermionic fields. After that we
study the quantization of this model and shows how $4D$ target space
supersymmetry follows.

\qquad The presentation of this paper is as follows: In section 2, we
develop a little bit our motivations and give preliminary results. In
section 3, we consider superspace representations of world sheet $\mathsf{n}%
=\left( 1,1\right) $ superalgebra and world line reductions. This super-
invariance allows to make an idea on the basis of our construction and
possible supersymmetric extension. In section 4, we give details on our
fermionic twistor like analog. First we comment the standard Penrose
construction for world line bosons. Then we build the pure fermionic model.
In section 5, we solve the constraint eqs and derive the quantum fermionic
twistor like field action.\ Section 6 is devoted to conclusion and
discussions.

\section{Motivations and preliminary results}

\qquad To make a general idea on our pure fermionic analogue of Penrose
twistor construction and on the method we will use to handle it, let us
start by recalling some useful features. We take also this opportunity to
give preliminary results that will be developed later.\newline
\qquad Consider a free world line bosonic field,
\begin{equation}
x^{m}\left( \tau \right) \sim x_{\alpha }^{\dot{\alpha}}\left( \tau \right)
,\qquad m=0,1,2,3,\qquad \alpha ,\dot{\alpha}=1,2
\end{equation}%
moving inside the real four dimension space $\mathbb{R}^{\mathrm{d},4-%
\mathrm{d}}$ with $d=2,3,4$. The field action $\mathcal{S}_{BT}$ describing
the dynamics of the massless particle reads as,
\begin{equation}
\mathcal{S}_{BT}\left[ x,p,e\right] =\int d\tau \left( p_{m}\partial _{\tau
}x^{m}-ep^{2}\right) ,  \label{seb}
\end{equation}%
where $p_{m}$ is the conjugate momentum of the world line field $x^{m}$ and $%
e=e\left( \tau \right) $ is the gauge field capturing the constraint eq
fixing the mass squared $p^{2}=M^{2}$ of the particle as,%
\begin{equation}
p^{2}=\eta ^{mn}p_{m}p_{n}=0.
\end{equation}%
To handle this equation, we rewrite the 4- component energy momentum vector $%
p_{m}$ like a hermitian $2\times 2$ matrix,
\begin{equation}
p_{m}\text{ }\sim \text{ }\left( \sigma _{m}\right) _{\alpha }^{\dot{\alpha}}%
\mathrm{p}_{\dot{\alpha}}^{\alpha }  \label{d1}
\end{equation}%
Here $\mathrm{p}_{\dot{\alpha}}^{\alpha }$ is a bi-spinor of the $SO\left(
\mathrm{d},4-\mathrm{d}\right) $ orthogonal group and the $\sigma _{m}$'s
are the usual Pauli matrices satisfying the Clifford algebra. The diagonal
metric $\eta _{mn}$ of $\mathbb{R}^{\mathrm{d},4-\mathrm{d}}$ can be also
expressed in terms of the product of the invariant spinor metrics $\epsilon
_{\alpha \beta }$ and $\epsilon _{\dot{\alpha}\dot{\beta}}$. Using dotted
and undotted spinor indices, we can re-express functions involving the
4-vector $p_{m}$ in terms of $2\times 2$ matrix $\mathrm{p}_{\dot{\alpha}%
}^{\alpha }$. For instance, the quadratic invariant $p^{2}$ is mapped to $%
\det \left( \mathrm{p}_{\dot{\alpha}}^{\alpha }\right) $ which can be also
rewritten as $\mathrm{Tr}\left( \mathrm{p}^{2}\right) $ with,%
\begin{eqnarray}
\mathrm{Tr}\left( \mathrm{p}^{2}\right)  &=&\epsilon _{\alpha \beta }\mathrm{%
p}_{\dot{\alpha}}^{\alpha }\mathrm{p}_{\dot{\beta}}^{\beta }\epsilon ^{\dot{%
\alpha}\dot{\beta}}=\mathrm{p}_{\dot{\alpha}}^{\alpha }.\mathrm{p}_{\alpha
}^{\dot{\alpha}}=\mathrm{p}_{\alpha }^{\dot{\alpha}}.\mathrm{p}_{\dot{\alpha}%
}^{\alpha },  \notag \\
\epsilon _{\alpha \beta }\epsilon ^{\beta \gamma } &=&\delta _{\alpha
}^{\gamma },\qquad \epsilon ^{\dot{\alpha}\dot{\beta}}\epsilon _{\dot{\beta}%
\dot{\gamma}}=\delta _{\dot{\gamma}}^{\dot{\alpha}}.
\end{eqnarray}%
From classical view, $p_{m}=$ $p_{m}\left( \tau \right) $ is a world line
field and the equation $\mathrm{Tr}\left( \mathrm{p}^{2}\right) =0$ admits a
remarkable solution given by,%
\begin{equation}
\mathrm{p}_{\dot{\alpha}}^{\alpha }=\lambda ^{\alpha }\widetilde{\lambda }_{%
\dot{\alpha}},\qquad \alpha ,\text{ }\dot{\alpha}=1,2,  \label{z1}
\end{equation}%
where the world line fields $\lambda ^{\alpha }=\lambda ^{\alpha }\left(
\tau \right) $ and $\widetilde{\lambda }_{\dot{\alpha}}=\widetilde{\lambda }%
_{\dot{\alpha}}\left( \tau \right) $ are $SO\left( \mathrm{d},4-\mathrm{d}%
\right) $ spinors with positive and negative chirality respectively.%
\begin{equation}
\lambda ^{\alpha }\sim \left( \frac{1}{2},0\right) ,\qquad \widetilde{%
\lambda }_{\dot{\alpha}}\sim \left( 0,\frac{1}{2}\right) .
\end{equation}%
By computing the invariant square $\mathrm{Tr}\left( \mathrm{p}^{2}\right) $
of the solution $\mathrm{p}_{\dot{\alpha}}^{\alpha }=\lambda ^{\alpha }%
\widetilde{\lambda }_{\dot{\alpha}}$, we see that the vanishing condition,
\begin{equation}
\mathrm{Tr}\left( \mathrm{p}^{2}\right) =\lambda ^{2}.\widetilde{\lambda }%
^{2}=0,
\end{equation}%
is ensured by requiring%
\begin{equation}
\lambda ^{2}=\epsilon _{\alpha \beta }\lambda ^{\alpha }\lambda ^{\beta
},\qquad \widetilde{\lambda }^{2}=\epsilon ^{\dot{\alpha}\dot{\beta}}%
\widetilde{\lambda }_{\dot{\alpha}}\widetilde{\lambda }_{\dot{\beta}}
\label{z2}
\end{equation}%
to vanish identically; i.e
\begin{equation}
\lambda ^{2}=0\qquad and/or\qquad \widetilde{\lambda }^{2}=0.
\end{equation}%
This requirement is naturally fulfilled in twistor analysis. Notice that the
vanishing property of the invariant squares of $\lambda ^{\alpha }$ and $%
\widetilde{\lambda }_{\dot{\alpha}}$ ($\lambda ^{2}=0$ and $\widetilde{%
\lambda }^{2}=0$) is due to the antisymmetry property of the dotted and
undotted metrics $\epsilon ^{\dot{\alpha}\dot{\beta}}=-\epsilon ^{\dot{\beta}%
\dot{\alpha}},$ $\epsilon _{\alpha \beta }=-\epsilon _{\beta \alpha }$ and
because of the commutativity of the Penrose variables \textrm{\cite{16,18}-%
\cite{1,27,28}},
\begin{equation}
\lambda ^{\alpha }\lambda ^{\beta }=\lambda ^{\beta }\lambda ^{\alpha
},\qquad \text{ and\ \qquad }\widetilde{\lambda }_{\dot{\alpha}}\widetilde{%
\lambda }_{\dot{\beta}}=\widetilde{\lambda }_{\dot{\beta}}\widetilde{\lambda
}_{\dot{\alpha}}.
\end{equation}%
However in looking carefully at this solution, one notes a set of remarkable
properties indicating that Penrose representation for $\mathrm{p}_{\dot{%
\alpha}}^{\alpha }=\lambda ^{\alpha }\widetilde{\lambda }_{\dot{\alpha}}$ is
just a part of a general picture which should deserve more attention. Below
we give three of them; they will be developed in much more details in
forthcoming sections.

\subsection{$\mathrm{p}_{\dot{\protect\alpha}}^{\protect\alpha }$: First
member of a field family $\left\{ {\protect\large \Pi }_{s\dot{\protect\alpha%
}}^{\protect\alpha }\text{ },\text{ s}\in \mathbb{Z}\text{ }\right\} $}

\qquad First note that to solve the condition,
\begin{equation}
\lambda ^{2}.\widetilde{\lambda }^{2}=0,
\end{equation}%
it is, a priori, enough to require $\lambda ^{2}=0$ whatever $\widetilde{%
\lambda }^{2}$ is or inversely. But this is not exactly what happens since
we have both $\lambda ^{2}=0$ and $\widetilde{\lambda }^{2}=0$. A careful
study shows that this degeneracy hides in fact a remarkable property which
can be used to go beyond the standard Penrose twistor construction for
bosons. In field theory, eq(\ref{z1}) can be viewed as just the leading
relation of a family of "conformal spin" $s$ world line field variables. A
class of these fields is given by the two following infinite sets:
\begin{equation}
\left( {\large \Pi }_{s}^{\prime }\right) _{\dot{\alpha}}^{\alpha }=\lambda
^{\alpha }\left( \widetilde{\lambda }_{s}\right) _{\dot{\alpha}},\qquad s=0,%
\text{ }\pm 2,\text{ }\pm 4,...,  \label{f1}
\end{equation}%
for world line bosons and by
\begin{equation}
\left( {\large \Pi }_{s}^{\prime }\right) _{\dot{\alpha}}^{\alpha }=\lambda
^{\alpha }\left( \widetilde{\zeta }_{s}\right) _{\dot{\alpha}},\qquad s=\pm
1,\text{ }\pm 3,\text{ }\pm 5,...,  \label{f2}
\end{equation}%
for world line fermions. In these relations $\lambda ^{\alpha }$ is as
before; but $\widetilde{\lambda }_{s\dot{\alpha}}$ and $\widetilde{\zeta }_{s%
\dot{\alpha}}$ are new fields with even and odd conformal spins $\frac{s}{2}$
respectively. As a first result we want to mention here is that, whatever
the value of $s$, all these $\left( {\large \Pi }_{s}^{\prime }\right) _{%
\dot{\alpha}}^{\alpha }$'s , $s\in \mathbb{Z}$, exhibit the property,
\begin{equation}
\mathrm{Tr}\left( {\large \Pi }_{s}^{\prime 2}\right) =\mathrm{\Pi }_{\dot{%
\alpha}}^{\alpha }.\mathrm{\Pi }_{\alpha }^{\dot{\alpha}}=0,
\end{equation}%
since,%
\begin{equation}
\mathrm{\Pi }_{\dot{\alpha}}^{\alpha }.\mathrm{\Pi }_{\alpha }^{\dot{\alpha}%
}=\epsilon _{\alpha \beta }\mathrm{\Pi }_{\dot{\alpha}}^{\alpha }\mathrm{\Pi
}_{\dot{\beta}}^{\beta }\epsilon ^{\dot{\alpha}\dot{\beta}}=\left\{
\begin{array}{c}
\lambda ^{2}.\widetilde{\lambda }_{s}^{2},\qquad s=0,\text{ }\pm 2,\text{ }%
\pm 4,... \\
\lambda ^{2}.\widetilde{\zeta }_{s}^{2},\qquad s=\pm 1,\text{ }\pm 3,\text{ }%
\pm 5,...%
\end{array}%
\right. ,
\end{equation}%
exactly as in eq(\ref{z1}). Nilpotency of $\mathrm{\Pi }_{\dot{\alpha}%
}^{\alpha }.\mathrm{\Pi }_{\alpha }^{\dot{\alpha}}$, the invariant square of
$\left( {\large \Pi }_{s}^{\prime }\right) _{\dot{\alpha}}^{\alpha }$'s is
then mainly due to
\begin{equation}
\lambda ^{2}=0,
\end{equation}%
since for the case of anticommuting $\widetilde{\zeta }_{s}$'s, the squares%
\begin{equation}
\left( \widetilde{\zeta }_{s}\right) ^{2}=\epsilon ^{\dot{\alpha}\dot{\beta}}%
\widetilde{\zeta }_{s\dot{\alpha}}\widetilde{\zeta }_{s\dot{\beta}}=\frac{1}{%
2}\epsilon ^{\dot{\alpha}\dot{\beta}}\left( \widetilde{\zeta }_{s\dot{\alpha}%
}\widetilde{\zeta }_{s\dot{\beta}}-\widetilde{\zeta }_{s\dot{\beta}}%
\widetilde{\zeta }_{s\dot{\alpha}}\right) \neq 0,
\end{equation}%
are non zero. Note that for the world line bosonic subset, we also have
\begin{equation}
\left( \widetilde{\lambda }_{s}\right) ^{2}=0.
\end{equation}%
and, in addition to $\mathrm{p}_{\dot{\alpha}}^{\alpha }=\lambda ^{\alpha }%
\widetilde{\lambda }_{0\dot{\alpha}}$, we shall also use the following
typical world line fields for our QFT model building and its dual twistor
like,%
\begin{equation}
{\large \Pi }_{\pm \dot{\alpha}}^{\prime \alpha }=\lambda ^{\alpha }%
\widetilde{\zeta }_{\pm \dot{\alpha}},\qquad \mathcal{P}_{\pm \pm \dot{\alpha%
}}^{\alpha }=\lambda ^{\alpha }\widetilde{\lambda }_{\pm \pm \dot{\alpha}}.
\end{equation}%
Here $\widetilde{\zeta }_{\pm \dot{\alpha}}$ and $\widetilde{\lambda }_{\pm
\pm \dot{\alpha}}$ are respectively anticommuting and commuting world line
fermions and bosons. Note moreover the following useful properties which can
help to get an overview on a possible stringy extension: \newline
(\textbf{a}) In general, the above result is not only valid for quantum
mechanics. These world line fields can be naturally promoted to world sheet
field variables where irreducible representations of the $SO\left(
2,R\right) $ rotation group of the real surface (plane) are one dimensional.
The charges $\pm q$ carried by the fields are nothing but the conformal
weights of 2D conformal invariance; see next subsection for technical
details.\newline
(\textbf{b}) Along with the set $\left( {\large \Pi }_{s}^{\prime }\right) _{%
\dot{\alpha}}^{\alpha }$ eqs(\ref{f1}-\ref{f2}), we have a second family $%
\left( {\large \Pi }_{s}^{\prime \prime }\right) _{\dot{\alpha}}^{\alpha }$
where the world line spin is carried by the undotted $SO\left( \mathrm{d},4-%
\mathrm{d}\right) $ spinors of target space. This family is given by,%
\begin{equation}
\left( {\large \Pi }_{s}^{\prime \prime }\right) _{\dot{\alpha}}^{\alpha
}=\left( \lambda _{s}\right) ^{\alpha }\text{ }\widetilde{\lambda }_{\dot{%
\alpha}},\qquad s=0,\text{ }\pm 2,\text{ }\pm 4,...,
\end{equation}%
for world line bosons and by
\begin{equation}
\left( {\large \Pi }_{s}^{\prime \prime }\right) _{\dot{\alpha}}^{\alpha
}=\left( \zeta _{s}\right) ^{\alpha }\text{ }\widetilde{\lambda }_{\dot{%
\alpha}},\qquad s=\pm 1,\text{ }\pm 3,\text{ }\pm 5,...,
\end{equation}%
for world line fermions. The vanishing condition $\mathrm{Tr}\left( {\large %
\Pi }_{s}^{\prime \prime 2}\right) =0$ is mainly ensured by the property
\begin{equation}
\widetilde{\lambda }^{2}=\widetilde{\lambda }_{\dot{\alpha}}\widetilde{%
\lambda }^{\dot{\alpha}}=0
\end{equation}%
for all values of $s$.\newline
(\textbf{c}) Both of the families ${\large \Pi }_{s\dot{\alpha}}^{\prime
\alpha }$ and ${\large \Pi }_{s\dot{\alpha}}^{\prime \prime \alpha }$ are in
fact two special subsets of the following general one,
\begin{equation}
\left( {\large \Pi }_{s}\right) _{\dot{\alpha}}^{\alpha }=\left(
f_{s_{1}}\right) ^{\alpha }\text{ }\widetilde{g}_{s_{2}\dot{\alpha}},\qquad
s=s_{1}+s_{2}.
\end{equation}%
Here the condition $\mathrm{Tr}\left( {\large \Pi }_{s}^{2}\right) =0$ is
not automatically fulfilled; it is solved whenever at least one of the two
world line (sheet) fields $f_{s_{1}}^{\alpha }$ and$\ \widetilde{g}_{s_{2}%
\dot{\alpha}}$ is a boson.

\subsection{${\protect\large \Pi }_{s\dot{\protect\alpha}}^{\protect\alpha %
}\left( \protect\tau \right) $ as boundary fields}

\qquad The above world line field variables, which we denote generically as $%
\phi =\phi \left( \tau \right) $, may be given various geometric
interpretations. For instance, they may be thought of as the boundary values,%
\begin{equation}
\phi \left( \tau \right) =\mathrm{\phi }\left( \sigma ^{-},\sigma
^{+}\right) |_{\partial \mathcal{M}}  \label{v}
\end{equation}%
of world sheet fields $\mathrm{\phi }\left( \sigma ^{-},\sigma ^{+}\right) $
of a $2d$ quantum (conformal) field theory (QFT$_{2}$). Here $\sigma ^{\pm }=%
\frac{1}{2}\left( \sigma _{0}\pm \sigma _{1}\right) $ are the variables
parameterizing the Riemann surface $\mathcal{M}$ with boundary $\partial
\mathcal{M}$ parameterized by $\tau $. In this view, the previous ${\large %
\Pi }_{s\dot{\alpha}}^{\alpha }\left( \tau \right) $'s are seen as the
boundary values associated with the world sheet fields,
\begin{equation}
\mathbf{\Pi }_{s\dot{\alpha}}^{\alpha }=\mathbf{\Pi }_{s\dot{\alpha}%
}^{\alpha }\left( \sigma ^{-},\sigma ^{+}\right) ,\qquad {\large \Pi }_{s%
\dot{\alpha}}^{\alpha }\left( \tau \right) =\mathbf{\Pi }_{s\dot{\alpha}%
}^{\alpha }\left( \sigma ^{-},\sigma ^{+}\right) |_{\partial \mathcal{M}},
\end{equation}%
and $s$ gets interpreted as the conformal spin, i.e \
\begin{equation}
\left( \sigma ^{+}\frac{\partial }{\partial \sigma ^{+}}-\sigma ^{-}\frac{%
\partial }{\partial \sigma ^{-}}\right) \mathbf{\Pi }_{s\dot{\alpha}%
}^{\alpha }\left( \sigma \right) \text{ }\sim \text{ }s\mathbf{\Pi }_{s\dot{%
\alpha}}^{\alpha }\left( \sigma \right) .
\end{equation}%
In section 3, we will give other examples by using reductions of world sheet
$\QTR{sl}{n}=\left( 1,1\right) $ superalgebra\footnote{%
As a convention of notations, we use the letter \textsf{n} to refer to
number of supercharges of world line (sheet) supersymmetry. Capital letter $%
\mathcal{N}$\ is deserved for target space supersymmetry.} with
supersymmetric covariant derivatives $D_{+}$ and $D_{-}$,%
\begin{equation}
D_{\mp }^{2}\text{ }\sim \text{ }\frac{\partial }{\partial \sigma ^{\pm }}
\end{equation}%
One of these examples is given by interpreting the world line variable $\tau
$ as the light cone coordinate,%
\begin{equation}
\sigma ^{+}=\frac{1}{2}\left( \sigma _{0}+\sigma _{1}\right) ,
\end{equation}%
parameterizing a left moving closed string or equivalently like $\sigma ^{-}=%
\frac{1}{2}\left( \sigma _{0}-\sigma _{1}\right) $ for right moving. An
other remarkable example, which we develop in present study, is given by the
association of the $\tau $ variable with an extra geometric dimension
realizing the central charge
\begin{equation}
\mathcal{Z}\text{ }\sim \text{ }D_{+}D_{-}+D_{-}D_{+}
\end{equation}%
of the $\QTR{sl}{n=}\left( 1,1\right) $ superalgebra; see also eqs(\ref{y1})
to fix the ideas. More precisely we have,%
\begin{equation}
\mathcal{Z}\phi \left( \tau \right) =\frac{\partial \phi \left( \tau \right)
}{\partial \tau }.
\end{equation}%
Before going into technical details, let us say few words about these tools
and where they will be used. This brings us to our next result.

\subsection{A pure fermionic model}

\qquad Part of this paper will be dealing with the derivation of a new pure
fermionic twistor like model. Instead of the massless particle described by
the world line field $x_{\dot{\alpha}}^{\alpha }\left( \tau \right) $ and
its conjugate momentum $\mathrm{p}_{\dot{\alpha}}^{\alpha }\left( \tau
\right) $ with $\mathrm{Tr}\left( \mathrm{p}^{2}\right) =0$, we consider
rather the world line fermions,
\begin{equation}
\Upsilon _{+\dot{\alpha}}^{\alpha }=\Upsilon _{+\dot{\alpha}}^{\alpha
}\left( \tau \right) ,\qquad \widetilde{\Upsilon }_{+\dot{\alpha}}^{\alpha }=%
\widetilde{\Upsilon }_{+\dot{\alpha}}^{\alpha }\left( \tau \right) ,
\end{equation}%
together with their conjugate momenta
\begin{equation}
{\Large \pi }_{-\dot{\alpha}}^{\alpha }={\Large \pi }_{-\dot{\alpha}%
}^{\alpha }\left( \tau \right) ,\qquad \widetilde{{\Large \pi }}_{-\dot{%
\alpha}}^{\alpha }=\widetilde{{\Large \pi }}_{-\dot{\alpha}}^{\alpha }\left(
\tau \right) .
\end{equation}%
Obviously this can be viewed as the simplest case one may consider.
Nevertheless, extensions involving much more fields is straightforward and
it turns out that they lead to extended target space supersymmetry. We sall
not study this issue in this paper; comments will given in the conclusion
and discussion section. \newline
We also have the twistor like splitting,
\begin{eqnarray}
\Upsilon _{+\dot{\alpha}}^{\alpha } &=&\lambda ^{\alpha }\widetilde{\zeta }%
_{+\dot{\alpha}},\qquad {\Large \pi }_{-\alpha }^{\dot{\alpha}}=\lambda
_{\alpha }\widetilde{\mathrm{\zeta }}_{-}^{\dot{\alpha}},  \notag \\
\widetilde{\Upsilon }_{+\dot{\alpha}}^{\alpha } &=&\mathrm{\zeta }%
_{+}^{\alpha }\widetilde{\lambda }_{\dot{\alpha}},\qquad \widetilde{{\Large %
\pi }}_{-\dot{\alpha}}^{\alpha }=\mathrm{\zeta }_{-}^{\alpha }\widetilde{%
\lambda }_{\dot{\alpha}}.  \label{gp}
\end{eqnarray}%
where $\lambda ^{\alpha }$ and $\widetilde{\lambda }_{\dot{\alpha}}$ are the
usual Penrose variable and $\mathrm{\zeta }_{\pm }^{\alpha }$ and $%
\widetilde{\zeta }_{\pm \dot{\alpha}}$ are anticommuting world line fermions,%
\begin{eqnarray}
\widetilde{\zeta }_{+\dot{\alpha}}\widetilde{\zeta }_{+\dot{\beta}}+%
\widetilde{\zeta }_{+\dot{\beta}}\widetilde{\zeta }_{+\dot{\alpha}} &=&0,
\notag \\
\widetilde{\zeta }_{-\dot{\alpha}}\widetilde{\zeta }_{+\dot{\beta}}+%
\widetilde{\zeta }_{+\dot{\beta}}\widetilde{\zeta }_{-\dot{\alpha}} &=&0,
\label{af} \\
\widetilde{\zeta }_{-\dot{\alpha}}\widetilde{\zeta }_{-\dot{\beta}}+%
\widetilde{\zeta }_{-\dot{\beta}}\widetilde{\zeta }_{-\dot{\alpha}} &=&0,
\notag
\end{eqnarray}%
and similar relations for undotted world line spinors. As it is known, the
conjugate momenta ${\Large \pi }_{\mp \alpha }^{\dot{\alpha}}$ for fermions
can be also imagined as just $\epsilon _{\alpha \beta }\epsilon ^{\dot{\alpha%
}\dot{\beta}}\Upsilon _{\mp \dot{\alpha}}^{\beta }$. By substituting the
expressions of $\Upsilon _{+\dot{\alpha}}^{\alpha }$ and ${\Large \pi }_{-%
\dot{\alpha}}^{\alpha }$ back into eqs(\ref{af}), one finds that they
satisfy as well the same anticommutation relations as the $\zeta $'s,%
\begin{eqnarray}
\Upsilon _{+\dot{\alpha}}^{\alpha }\Upsilon _{+\dot{\beta}}^{\beta
}+\Upsilon _{+\dot{\beta}}^{\beta }\Upsilon _{+\dot{\alpha}}^{\alpha } &=&0,
\notag \\
{\Large \pi }_{-\alpha }^{\dot{\alpha}}\Upsilon _{+\dot{\beta}}^{\beta
}+\Upsilon _{+\dot{\beta}}^{\beta }{\Large \pi }_{-\alpha }^{\dot{\alpha}}
&=&0, \\
{\Large \pi }_{-\alpha }^{\dot{\alpha}}{\Large \pi }_{-\beta }^{\dot{\beta}}+%
{\Large \pi }_{-\beta }^{\dot{\beta}}{\Large \pi }_{-\alpha }^{\dot{\alpha}}
&=&0.  \notag
\end{eqnarray}%
Similar relations are also valid for $\widetilde{\Upsilon }$ and $\widetilde{%
\Pi }$. By taking the trace of these relations and their tildes, we obtain
the following
\begin{eqnarray}
\mathrm{Tr}\left( \Upsilon _{+}^{2}\right) &=&0,\qquad \mathrm{Tr}\left(
\widetilde{\Upsilon }_{+}^{2}\right) =0  \notag \\
\mathrm{Tr}\left( {\Large \pi }_{-}^{2}\right) &=&0,\qquad \mathrm{Tr}\left(
\widetilde{{\Large \pi }}_{-}^{2}\right) =0  \label{ac} \\
\mathrm{Tr}\left( {\Large \pi }_{-}\Upsilon _{+}\right) &=&0,\qquad \mathrm{%
Tr}\left( \widetilde{{\Large \pi }}_{-}\widetilde{\Upsilon }_{+}\right) =0.
\notag
\end{eqnarray}%
they are nothing but the field property $\mathrm{Tr}\left( {\large \Upsilon }%
_{\pm }^{2}\right) =0$ and partners derived in sub-section 2.1. The pure
fermionic world line field action describing the dynamics of these fields is
given,%
\begin{eqnarray}
\mathcal{S}_{F}\left[ \Upsilon ,{\Large \pi },\widetilde{\Upsilon },%
\widetilde{{\Large \pi }}\right] &=&\int_{\mathbb{R}}d\tau \mathrm{Tr}\left(
{\Large \pi }_{-\dot{\alpha}}^{\alpha }\frac{\partial }{\partial \tau }%
\Upsilon _{+\alpha }^{\dot{\alpha}}+\widetilde{{\Large \pi }}_{-\dot{\alpha}%
}^{\alpha }\frac{\partial }{\partial \tau }\widetilde{\Upsilon }_{+\alpha }^{%
\dot{\alpha}}\right)  \notag \\
&&-\int_{\mathbb{R}}d\tau \mathrm{Tr}\left[ \mathrm{\eta }_{--}\left(
\Upsilon _{+}^{2}\right) +\mathrm{\eta }_{++}\left( {\Large \pi }%
_{-}^{2}\right) +\mathrm{\eta }_{+-}\left( {\Large \pi }_{-}\Upsilon
_{+}\right) \right]  \label{pfa} \\
&&-\int_{\mathbb{R}}d\tau \mathrm{Tr}\left[ \widetilde{\mathrm{\eta }}%
_{--}\left( \widetilde{\Upsilon }_{+}^{2}\right) +\widetilde{\mathrm{\eta }}%
_{++}\left( \widetilde{{\Large \pi }}_{-}^{2}\right) +\widetilde{\mathrm{%
\eta }}_{+-}\left( \widetilde{{\Large \pi }}_{-}\widetilde{\Upsilon }%
_{+}\right) \right] .  \notag
\end{eqnarray}%
where $\mathrm{\eta }_{\pm \pm }$, $\mathrm{\eta }_{+-}$ and $\widetilde{%
\mathrm{\eta }}_{\pm \pm }$, $\widetilde{\mathrm{\eta }}_{+-}$ are auxiliary
world line gauge fields capturing the three constraint eqs(\ref{ac}). We
will see later; in particular for the case of real world line fermions, that
these field constraints are captured by\ a $SP\left( 2,R\right) \times
SP\left( 2,R\right) $ gauge symmetry.

\subsection{World line (sheet) \& target space supersymmetries}

\qquad In above field action, one notes that the solution (\ref{gp}) of the
constraint eqs(\ref{ac}), which can be defined also as,
\begin{equation}
\frac{\delta \mathcal{S}_{F}}{\delta \mathrm{\eta }_{rs}}=0,\qquad \frac{%
\delta \mathcal{S}_{F}}{\delta \widetilde{\mathrm{\eta }}_{rs}}=0,\qquad
r,s=\pm
\end{equation}%
involves two kinds of world line field pairs: (\textbf{i}) the world line
bosons
\begin{equation}
\lambda ^{\alpha },\qquad \widetilde{\lambda }_{\dot{\alpha}},
\end{equation}%
and (\textbf{ii}) the fermionic partners%
\begin{equation}
\zeta _{+}^{\alpha },\qquad \widetilde{\zeta }_{+\dot{\alpha}}.
\end{equation}%
If supplementing these twistor like variables by their conjugate momenta
namely%
\begin{equation}
\mathrm{\mu }^{\alpha },\qquad \widetilde{\mathrm{\mu }}_{\dot{\alpha}%
},\qquad \mathrm{\zeta }_{-}^{\alpha },\qquad \widetilde{\mathrm{\zeta }}_{-%
\dot{\alpha}},
\end{equation}%
one sees that these fields altogether form world line supersymmetric
multiplets as shown below,%
\begin{equation}
\left( \lambda ^{\alpha },\zeta _{+}^{\alpha }\right) ,\qquad \left(
\widetilde{\lambda }_{\dot{\alpha}},\widetilde{\zeta }_{+\dot{\alpha}%
}\right) ,\qquad \left( \mathrm{\mu }^{\alpha },\mathrm{\zeta }_{-}^{\alpha
}\right) ,\qquad \left( \widetilde{\mathrm{\mu }}_{\dot{\alpha}},\widetilde{%
\mathrm{\zeta }}_{-\dot{\alpha}}\right) .
\end{equation}%
This property shows that the field action (\ref{pfa}) can be made
supersymmetric. In this subsection, we first give preliminary results on
world line (sheet) extensions of our model. Then we give the link between
the pure fermionic model and target space supersymmetry.

\subsubsection{World line supersymmetry}

In standard Penrose construction, one uses the world line twistor field
variables,%
\begin{equation}
\lambda ^{\alpha }=\lambda ^{\alpha }\left( \tau \right) ,\text{ \ \qquad }%
\mathrm{\mu }^{\dot{\alpha}}=\mathrm{\mu }^{\dot{\alpha}}\left( \tau \right)
,
\end{equation}%
together with their compagnons with opposite $SO\left( \mathrm{d},4-\mathrm{d%
}\right) $ chiralities, $\mathrm{d}=2,3,4$,%
\begin{equation}
\widetilde{\mathrm{\mu }}^{\alpha }=\widetilde{\mathrm{\mu }}^{\alpha
}\left( \tau \right) ,\text{ \ \qquad }\widetilde{\lambda }_{\dot{\alpha}}=%
\widetilde{\lambda }_{\dot{\alpha}}\left( \tau \right) .  \label{se}
\end{equation}%
As the intrinsic properties of these spinors depends on the value of $%
\mathrm{d}$, the structure of the underlying QFTs is sensitive to these
features. For the target space $\mathbb{R}^{\left( \mathrm{2,2}\right) }$
for instance, the homogeneity group $SO\left( \mathrm{2},2,R\right) $ is
locally isomorphic $SL\left( \mathrm{2},R\right) \times \widetilde{SL}\left(
\mathrm{2},R\right) $ and the above component field spinors are basically
real; a specific property making the $\mathbb{R}^{\left( \mathrm{2,2}\right)
}$ analysis a little bit subtle. However, for the spaces $\mathbb{R}^{\left(
\mathrm{1,3}\right) }$ and $\mathbb{R}^{\mathrm{4}}$, the corresponding
rotation groups $SO\left( \mathrm{1},3,R\right) $ and $SO\left( \mathrm{4}%
,R\right) $ are respectively isomorphic to $SL\left( \mathrm{2},C\right) $
and $SU\left( \mathrm{2},C\right) \times SU\left( \mathrm{2},C\right) $ and
the fundamental spinors are complex. For the example of the $SL\left(
\mathrm{2},C\right) $ case, the sector with $\widetilde{\mathrm{\mu }}%
^{\alpha }$ and $\widetilde{\lambda }_{\dot{\alpha}}$ eq(\ref{se}) is just
the complex conjugate of $\mathrm{\mu }^{\dot{\alpha}}$ and $\lambda
^{\alpha }$, i.e,
\begin{equation}
\overline{\mathrm{\mu }}^{\alpha }=\overline{\left( \mathrm{\mu }_{\dot{%
\alpha}}\right) },\text{ \ \qquad }\overline{\lambda }_{\dot{\alpha}}=%
\overline{\left( \lambda ^{\alpha }\right) }.
\end{equation}%
This complex nature of the twistor variables plays a crucial role in the
algebraic geometry analysis of twistor space where holomorphic and
antiholomorphic sectors factorise. \newline
In general, the field $\mathrm{\mu }^{\dot{\alpha}}=\frac{\delta S_{T}}{%
\delta \partial _{\tau }\widetilde{\lambda }_{\dot{\alpha}}}$ (resp. $%
\widetilde{\mathrm{\mu }}_{\alpha }=\frac{\delta S_{T}}{\delta \partial
_{\tau }\lambda ^{\alpha }}$) is the conjugate momentum of $\lambda ^{\alpha
}$ (resp. $\widetilde{\lambda }_{\dot{\alpha}}$),
\begin{equation}
\left\{ \widetilde{\lambda }_{\dot{\alpha}},\mathrm{\mu }^{\dot{\beta}%
}\right\} _{TPB}=\delta _{\dot{\alpha}}^{\dot{\beta}},
\end{equation}%
where $\left\{ f,\mathrm{g}\right\} _{TPB}$ stands for Poisson bracket in
twistor space. To make contact with $\mathbb{R}^{\left( \mathrm{d,4-d}%
\right) }$ geometry, $\lambda ^{\alpha }$ and $\mathrm{\mu }^{\dot{\alpha}}$%
\ should be compared with the usual phase space variables,%
\begin{equation}
x^{m}\sim x_{\alpha }^{\dot{\alpha}}\left( \tau \right) ,\qquad p_{m}\sim p_{%
\dot{\alpha}}^{\alpha }\left( \tau \right)  \label{bo}
\end{equation}%
with $\left\{ x^{m},p_{n}\right\} _{MPB}=\delta _{n}^{m}$ being the Poisson
bracket. \newline
In the twistor like fermions we are considering here, we have,%
\begin{equation}
\widetilde{\zeta }_{+\dot{\alpha}}=\widetilde{\zeta }_{+\dot{\alpha}}\left(
\tau \right) ,\text{ \ }\qquad \mathrm{\xi }_{-}^{\dot{\alpha}}=\mathrm{\xi }%
_{-}^{\dot{\alpha}}\left( \tau \right) ,
\end{equation}%
with $\mathrm{\xi }_{-}^{\dot{\alpha}}=\frac{\delta \mathcal{S}_{FT}}{\delta
\left( \partial _{\tau }\widetilde{\zeta }_{+\dot{\alpha}}\right) }$ and
where $\mathcal{S}_{FT}$ is the field action of the twistor like variables.
We also have
\begin{equation}
\left\{ \widetilde{\zeta }_{+\dot{\alpha}},\mathrm{\xi }_{-}^{\dot{\beta}%
}\right\} _{TGPB}=\delta _{\dot{\alpha}}^{\dot{\beta}}
\end{equation}%
where now $\left\{ f_{+},\mathrm{g}_{-}\right\} _{TGPB}$ is the graded
Poisson bracket in twistor space. These fermionic variables $\widetilde{%
\zeta }_{+\dot{\alpha}}$ and $\mathrm{\xi }_{-}^{\dot{\alpha}}$ should be
thought of as the twistor dual to the world line fermions in 4- vector of $%
SO\left( \mathrm{d},4-\mathrm{d}\right) $,
\begin{eqnarray}
\Upsilon _{+}^{m} &\sim &\Upsilon _{+\alpha }^{\dot{\alpha}}\left( \tau
\right) ,\qquad {\large \pi }_{-m}\sim {\large \pi }_{-\dot{\alpha}}^{\alpha
}\left( \tau \right)  \notag \\
\widetilde{\Upsilon }_{+}^{m} &\sim &\widetilde{\Upsilon }_{+\alpha }^{\dot{%
\alpha}}\left( \tau \right) ,\qquad \widetilde{{\large \pi }}_{-m}\sim
\widetilde{{\large \pi }}_{-\dot{\alpha}}^{\alpha }\left( \tau \right)
\label{fe}
\end{eqnarray}%
If forgetting for a while about the world line variable $\tau $, one remarks
that the combination of eqs(\ref{bo}-\ref{fe}) describe pairs $\left(
x,\Upsilon _{+},\widetilde{\Upsilon }_{+}\right) $ and $\left( p,{\large \pi
}_{-},\widetilde{{\large \pi }}_{-}\right) $, which can be viewed as world
line supermultiplets,%
\begin{equation}
\left( x^{m},\Upsilon _{+}^{m},\widetilde{\Upsilon }_{+}^{m}\right) ,\qquad
\left( p_{m},{\large \pi }_{-m},\widetilde{{\large \pi }}_{-m}\right)
,\qquad m=0,1,2,3.
\end{equation}%
A similar conclusion may be done regarding the supermultiplets built out of
twistor like variables $\lambda ^{\alpha },$ $\widetilde{\zeta }_{+\dot{%
\alpha}},$ $\mathrm{\mu }^{\dot{\beta}}\mathrm{,\xi }_{-}^{\dot{\beta}}$ and
their partners involving the spinors with opposite chirality. Using the
representations,
\begin{eqnarray}
p_{m} &\sim &\mathrm{p}_{\dot{\alpha}}^{\alpha }=\lambda ^{\alpha }%
\widetilde{\lambda }_{\dot{\alpha}},  \notag \\
{\large \pi }_{-m} &\sim &{\large \pi }_{-\dot{\alpha}}^{\alpha }=\zeta
_{-}^{\alpha }\widetilde{\lambda }_{\dot{\alpha}}, \\
\widetilde{{\large \pi }}_{-m} &\sim &\widetilde{{\large \pi }}_{-\dot{\alpha%
}}^{\alpha }=\lambda ^{\alpha }\widetilde{\zeta }_{-\dot{\alpha}},  \notag
\end{eqnarray}%
it is not difficult to check that world line supersymmetric transformations,%
\begin{eqnarray}
\delta \mathrm{p}_{\dot{\alpha}}^{\alpha } &\sim &\varepsilon _{+}{\large %
\pi }_{-\dot{\alpha}}^{\alpha }+\widetilde{\varepsilon }_{+}\widetilde{%
{\large \pi }}_{-\dot{\alpha}}^{\alpha },  \notag \\
\delta {\large \pi }_{-\dot{\alpha}}^{\alpha } &\sim &\varepsilon _{+}\frac{d%
\mathrm{p}_{\dot{\alpha}}^{\alpha }}{d\tau },\qquad \delta \widetilde{%
{\large \pi }}_{-\dot{\alpha}}^{\alpha }\sim \widetilde{\varepsilon }_{+}%
\frac{d\mathrm{p}_{\dot{\alpha}}^{\alpha }}{d\tau }
\end{eqnarray}%
translate as well to the twistor like space. By substitution, we get $\delta
\mathrm{p}_{\dot{\alpha}}^{\alpha }=$ $\left( \delta \lambda ^{\alpha
}\right) \widetilde{\lambda }_{\dot{\alpha}}+$ $\lambda ^{\alpha }\left(
\delta \widetilde{\lambda }_{\dot{\alpha}}\right) $ which should be
identified with $\left( \varepsilon _{+}\zeta _{-}^{\alpha }\right)
\widetilde{\lambda }_{\dot{\alpha}}+\lambda ^{\alpha }\left( \widetilde{%
\varepsilon }_{+}\widetilde{\zeta }_{-\dot{\alpha}}\right) $. Thus we have,%
\begin{eqnarray}
\delta \lambda ^{\alpha } &\sim &\varepsilon _{+}\zeta _{-}^{\alpha },\qquad
\delta \widetilde{\lambda }_{\dot{\alpha}}\sim \widetilde{\varepsilon }_{+}%
\widetilde{\zeta }_{-\dot{\alpha}},  \notag \\
\delta \zeta _{-}^{\alpha } &\sim &\varepsilon _{+}\frac{d\lambda ^{\alpha }%
}{d\tau },\qquad \delta \widetilde{\zeta }_{-\dot{\alpha}}\sim \widetilde{%
\varepsilon }_{+}\frac{d\widetilde{\lambda }_{\dot{\alpha}}}{d\tau }.
\end{eqnarray}%
The pure fermionic construction to be developed in this paper may be then
viewed as a sub-sector of the odd part of a super-world line extension of
Penrose model.

\subsubsection{World sheet supersymmetry}

\qquad Following the view given by eq(\ref{v}), the above super-world line
field model can embedded in turn in world sheet $\QTR{sl}{n}=\left(
1,1\right) $ supersymmetric field theory with target space $\mathbb{R}%
^{\left( \mathrm{d},\mathrm{4-d}\right) }$. There, generic fields $\phi $
are functions of the world sheet variables $\sigma =\left( \sigma
^{-},\sigma ^{+}\right) $. The field variables,%
\begin{eqnarray}
x_{\dot{\alpha}}^{\alpha } &=&x_{\dot{\alpha}}^{\alpha }\left( \sigma
\right) ,\qquad \Upsilon _{\pm \dot{\alpha}}^{\alpha }=\Upsilon _{\pm \dot{%
\alpha}}^{\alpha }\left( \sigma \right) ,  \notag \\
{\large p}_{\dot{\alpha}}^{\alpha } &=&{\large p}_{\dot{\alpha}}^{\alpha
}\left( \sigma \right) ,\qquad {\large \pi }_{\pm \dot{\alpha}}^{\alpha }=%
{\large \pi }_{\pm \dot{\alpha}}^{\alpha }\left( \sigma \right) ,
\end{eqnarray}%
and the corresponding twistor like variables,
\begin{eqnarray}
\lambda ^{\alpha } &=&\lambda ^{\alpha }\left( \sigma \right) ,\qquad \zeta
_{\pm }^{\alpha }=\zeta _{\pm }^{\alpha }\left( \sigma \right) ,  \notag \\
\mathrm{\mu }^{\dot{\alpha}} &=&\mathrm{\mu }^{\dot{\alpha}}\left( \sigma
\right) ,\qquad \mathrm{\xi }_{\mp }^{\dot{\alpha}}=\mathrm{\xi }_{\mp }^{%
\dot{\alpha}}\left( \sigma \right) ,  \notag \\
\widetilde{\lambda }_{\dot{\alpha}} &=&\widetilde{\lambda }_{\dot{\alpha}%
}\left( \sigma \right) ,\qquad \widetilde{\zeta }_{\pm \dot{\alpha}}=%
\widetilde{\zeta }_{\pm \dot{\alpha}}\left( \sigma \right) \\
\widetilde{\mathrm{\mu }}_{\alpha } &=&\widetilde{\mathrm{\mu }}_{\alpha
}\left( \sigma \right) ,\qquad \widetilde{\mathrm{\xi }}_{\mp \alpha }=%
\widetilde{\mathrm{\xi }}_{\mp \alpha }\left( \sigma \right) ,  \notag
\end{eqnarray}%
form world sheet supermultiplets. We also have the following graded Poisson
brackets,%
\begin{eqnarray}
\left\{ x_{\dot{\alpha}}^{\alpha }\left( \sigma \right) ,{\large p}_{\beta
}^{\dot{\beta}}\left( \sigma ^{\prime }\right) \right\} &=&\delta _{\beta
}^{\alpha }\delta _{\dot{\alpha}}^{\dot{\beta}}\delta \left( \sigma -\sigma
^{\prime }\right)  \notag \\
\left\{ \lambda ^{\alpha }\left( \sigma \right) ,\widetilde{\mathrm{\mu }}%
_{\beta }\left( \sigma ^{\prime }\right) \right\} &=&\delta _{\beta
}^{\alpha }\delta \left( \sigma -\sigma ^{\prime }\right)  \notag \\
\left\{ \widetilde{\lambda }_{\dot{\alpha}}\left( \sigma \right) ,\mathrm{%
\mu }^{\dot{\beta}}\left( \sigma ^{\prime }\right) \right\} &=&\delta _{\dot{%
\alpha}}^{\dot{\beta}}\delta \left( \sigma -\sigma ^{\prime }\right) \\
\left\{ \zeta _{\pm }^{\alpha }\left( \sigma \right) ,\widetilde{\mathrm{\xi
}}_{\mp \beta }\left( \sigma ^{\prime }\right) \right\} &=&\delta _{\beta
}^{\alpha }\delta \left( \sigma -\sigma ^{\prime }\right)  \notag \\
\left\{ \widetilde{\zeta }_{\pm \dot{\alpha}}\left( \sigma \right) ,\mathrm{%
\xi }_{\mp }^{\dot{\beta}}\left( \sigma ^{\prime }\right) \right\} &=&\delta
_{\dot{\alpha}}^{\dot{\beta}}\delta \left( \sigma -\sigma ^{\prime }\right) ,
\notag
\end{eqnarray}%
and all remaining others are zero. Typical world sheet supersymmetric
transformations are given by,%
\begin{eqnarray}
\delta _{\varepsilon }x_{\dot{\alpha}}^{\alpha }\left( \sigma \right) \text{
} &\sim &\text{ }i\varepsilon _{+}\Upsilon _{-\dot{\alpha}}^{\alpha }\left(
\sigma \right) +i\varepsilon _{-}\Upsilon _{+\dot{\alpha}}^{\alpha }\left(
\sigma \right)  \notag \\
\delta _{\varepsilon }\Upsilon _{\pm \dot{\alpha}}^{\alpha }\left( \sigma
\right) \text{ } &\sim &\text{ }\varepsilon _{\mp }\partial _{\pm \pm }x_{%
\dot{\alpha}}^{\alpha }\left( \sigma \right) .
\end{eqnarray}%
$\varepsilon _{\mp }$ are the parameters of the $\QTR{sl}{n}=\left(
1,1\right) $ superalgebra to be studied in next section. Similar
transformations may be written down for $\left( {\large p}_{\dot{\alpha}%
}^{\alpha }\left( \sigma \right) ,{\large \pi }_{-\dot{\alpha}}^{\alpha
}\left( \sigma \right) \right) $ and the corresponding twistor like ones.

\subsubsection{Target space supersymmetry}

\qquad The result we give here is that the quantum spectrum of the pure
fermionic with field action $\mathcal{S}_{F}\left[ \Upsilon ,{\Large \pi },%
\widetilde{\Upsilon },\widetilde{{\Large \pi }}\right] $ eqs(\ref{pfa}) has $%
4D$ target space supersymmetry. To derive $4D$ $\mathcal{N}=1$ target space
supersymmetry, consider the following particular twistor like fermionic
world line field action,%
\begin{eqnarray}
\mathcal{S}_{FT}\text{ } &\sim &\text{ }\int d\tau \left[ \widetilde{\mathrm{%
\mu }}_{\alpha }\partial _{\tau }\lambda ^{\alpha }-\mathrm{\xi }_{-}^{\dot{%
\alpha}}\partial _{\tau }\widetilde{\zeta }_{+\dot{\alpha}}\right]  \notag \\
&&-\int d\tau \left[ \mathrm{\mu }^{\dot{\alpha}}\partial _{\tau }\widetilde{%
\lambda }_{\dot{\alpha}}-\widetilde{\mathrm{\xi }}_{-\alpha }\partial _{\tau
}\zeta _{+}^{\alpha }\right] .  \label{tr}
\end{eqnarray}%
This action has a set of symmetries to be studied in section 5 and lead to $%
\mathcal{N}=1$ supersymmetric quantum spectrum in target space coordinated
by $\left\{ \lambda ^{\alpha },\mathrm{\mu }^{\dot{\alpha}}\right\} $. Eq(%
\ref{tr}) involves two world line fermions $\zeta _{-}^{\alpha }$ and $%
\mathrm{\mu }^{\dot{\alpha}}$ together with their conjugate momenta; more
world line spinors leads to $4D$ target space extended supersymmetry.
\newline
Target space supersymmetry is obtained by quantizing the system. This is
achieved by promoting the graded Poisson brackets the phase space variables,%
\begin{eqnarray}
\left\{ \widetilde{\lambda }_{\dot{\alpha}},\mathrm{\mu }^{\dot{\beta}%
}\right\} _{GPB}\text{ } &\sim &\text{ }\delta _{\dot{\alpha}}^{\dot{\beta}%
},\qquad \left\{ \lambda ^{\alpha },\widetilde{\mathrm{\mu }}_{\beta
}\right\} _{GPB}\text{ }\sim \text{ }\delta _{\beta }^{\alpha },  \notag \\
\left\{ \widetilde{\zeta }_{+\dot{\alpha}},\mathrm{\xi }_{-}^{\dot{\beta}%
}\right\} _{GPB}\text{ } &\sim &\text{ }\delta _{\dot{\alpha}}^{\dot{\beta}%
},\qquad \left\{ \zeta _{+}^{\alpha },\widetilde{\mathrm{\xi }}_{-\beta
}\right\} _{GPB}\text{ }\sim \text{ }\delta _{\beta }^{\alpha },
\end{eqnarray}%
into graded canonical commutation relations.
\begin{eqnarray}
\left[ \widetilde{\lambda }_{\dot{\alpha}},\mathrm{\mu }^{\dot{\beta}}\right]
_{-}\text{ } &\sim &\text{ }\delta _{\dot{\alpha}}^{\dot{\beta}},\qquad %
\left[ \lambda ^{\alpha },\widetilde{\mathrm{\mu }}_{\beta }\right] _{-}%
\text{ }\sim \text{ }\delta _{\beta }^{\alpha },  \notag \\
\left[ \widetilde{\zeta }_{+\dot{\alpha}},\mathrm{\xi }_{-}^{\dot{\beta}}%
\right] _{+}\text{ } &\sim &\text{ }\delta _{\dot{\alpha}}^{\dot{\beta}%
},\qquad \left[ \zeta _{+}^{\alpha },\widetilde{\mathrm{\xi }}_{-\beta }%
\right] _{+}\text{ }\sim \text{ }\delta _{\beta }^{\alpha }.
\end{eqnarray}%
Here $\left[ \text{\ }_{,}\text{ }\right] _{-}$ and $\left[ \text{\ }_{,}%
\text{ }\right] _{+}$\ (to avoid confusion with graded Poisson bracket $%
\left\{ \text{ },\text{ }\right\} _{GPB}$ ) stand for commutator and
anticommutator respectively. Then solve these relations like
\begin{eqnarray}
\mathrm{\mu }^{\dot{\alpha}} &\sim &\frac{\partial }{\partial \widetilde{%
\lambda }_{\dot{\alpha}}},\qquad \widetilde{\mathrm{\mu }}_{\alpha }\sim
\frac{\partial }{\partial \lambda ^{\alpha }},  \notag \\
\mathrm{\xi }_{-}^{\dot{\alpha}} &\sim &\frac{\partial }{\partial \widetilde{%
\zeta }_{+\dot{\alpha}}},\qquad \widetilde{\mathrm{\xi }}_{+\alpha }\sim
\frac{\partial }{\partial \zeta _{-}^{\alpha }}.
\end{eqnarray}%
In this representation, the conjugate momenta are then given by differential
operators which act on the wave functions $\Phi $ living on a graded space
as shown below,
\begin{equation}
\Phi =\Phi \left( \lambda ^{\alpha },\text{ }\mathrm{\mu }^{\dot{\alpha}}|%
\text{ }\zeta _{+}^{\alpha },\text{ }\mathrm{\xi }_{-}^{\dot{\alpha}}\right)
.
\end{equation}%
Since $\zeta _{+}^{\alpha }$ and $\mathrm{\xi }_{-}^{\dot{\alpha}}$ are
fermions, the waves $\Phi $ can be thought of as superfields on the
superspace $\mathbb{R}^{\left( \mathrm{d,4-d}|4\right) }$; and moreover can
be expanded as usual into a finite series of power of $\zeta _{-}^{\alpha }$
and $\mathrm{\xi }_{-}^{\dot{\alpha}}$. These superfields define just off
shell representations of the four dimensional $\mathcal{N}=1$ supersymmetry,%
\begin{eqnarray}
\left\{ \mathrm{D}_{+\alpha },\mathrm{D}_{-}^{\dot{\alpha}}\right\} \text{ }
&=&2\text{ \textrm{P}}_{\alpha }^{\dot{\alpha}},  \notag \\
\left[ \text{\textrm{P}}_{\alpha }^{\dot{\alpha}},\mathrm{D}_{+\alpha }%
\right] \text{ } &=&\text{ }\left[ \text{\textrm{P}}_{\alpha }^{\dot{\alpha}%
},\mathrm{D}_{-}^{\dot{\alpha}}\right] \text{ }=0.  \label{spa}
\end{eqnarray}%
In twistor like variables, this superalgebra can be realized as,
\begin{eqnarray}
\mathrm{D}_{+\dot{\alpha}} &=&\frac{\partial }{\partial \mathrm{\xi }_{-}^{%
\dot{\alpha}}}+\zeta _{+}^{\alpha }\lambda _{\alpha }\frac{\partial }{%
\partial \mathrm{\mu }^{\dot{\alpha}}},  \notag \\
\mathrm{D}_{-\alpha } &=&\frac{\partial }{\partial \zeta _{+}^{\alpha }}+%
\mathrm{\xi }_{-}^{\dot{\alpha}}\mathrm{\mu }_{\dot{\alpha}}\frac{\partial }{%
\partial \lambda ^{\alpha }}, \\
\text{\textrm{P}}_{\alpha }^{\dot{\alpha}} &=&\lambda _{\alpha }\frac{%
\partial }{\partial \mathrm{\mu }_{\dot{\alpha}}}+\mathrm{\mu }^{\dot{\alpha}%
}\frac{\partial }{\partial \lambda ^{\alpha }}.  \notag
\end{eqnarray}%
To fix the ideas, let us focus on $\mathbb{R}^{\left( 1,3|4\right) }$
superspace and give the example of $4D$ $\mathcal{N}=1$ chiral multiplet;
similar analysis can be done for gauge multiplet. In this case, $\widetilde{%
\lambda }_{\dot{\alpha}}=\overline{\lambda }_{\dot{\alpha}}$, $\widetilde{%
\zeta }_{+}^{\dot{\alpha}}=\overline{\zeta }_{+}^{\dot{\alpha}}$ and eq(\ref%
{spa}) has a chiral representation. The superfield $\Phi $ has no dependence
in $\overline{\mathrm{\xi }}_{-}^{\dot{\alpha}}$ and its expansion reads as
follows,
\begin{equation}
\Psi \left( \lambda ,\mathrm{\mu }|\text{ }\zeta _{+}^{\alpha }\right) =\phi
\left( \lambda ,\mathrm{\mu }\right) +\zeta _{+}^{\alpha }\psi _{-\alpha
}\left( \lambda ,\mathrm{\mu }\right) +\zeta _{+}^{2}F_{--}\left( \lambda ,%
\mathrm{\mu }\right) .  \label{ex}
\end{equation}%
The component fields $\left( \phi ,\psi _{\alpha },F_{--}\right) $ are
nothing but the scalar multiplet of four dimensional $\mathcal{N}=1$
superalgebra(\ref{spa}). For other supersymmetric representations and
explicit details for the target spaces $\mathbb{R}^{\left( \mathrm{d,4-d}%
\right) }$, $d=2,3,4$, see section 5 and eqs(\ref{50},\ref{51},\ref{h}).

\section{Embedding world line models in 2D superalgebra}

\qquad Because of geometric interpretation purpose and for later use, we
start this section by discussing briefly the general superspace realization
of the world sheet $\mathsf{n}=\left( 1,1\right) $ supersymmetric algebra.
Then we consider particular cases associated with some special subsymmetries
of this $\mathbb{Z}_{2}$ graded algebra.\ This helps us to derive the
symmetry properties of the world line fields of the twistor like fermion
model.

\qquad Roughly, the $\mathsf{n}=\left( 1,1\right) $ superalgebra lives on
Riemann surfaces $\mathcal{M}$ (string world sheet) and can, roughly, be
defined on a local chart of $\mathcal{M}$ as,%
\begin{eqnarray}
\{D_{+},D_{+}\} &=&-2\mathrm{a}\partial _{++},  \notag \\
\{D_{-},D_{-}\} &=&-2\mathrm{b}\partial _{--}  \notag \\
\{D_{\pm },D_{\mp }\} &=&-2\mathrm{c}\mathcal{Z}  \label{y1} \\
\left[ \partial _{\pm \pm },Q_{\pm }\right] &=&\left[ \partial _{\pm \pm },%
\mathcal{Z}\right] =\left[ \mathcal{Z},Q_{\pm }\right] =0,  \notag
\end{eqnarray}%
In these relations, the $D_{\pm }$'s are the usual supersymmetric covariant
derivatives of the superalgebra and $\mathcal{Z}$ is its unique central
charge. The latter is ignored in almost all simple applications; but in a
complete analysis it should be taken into account as plays a crucial role in
the analysis of BPS states and non perturbative dynamics. The extra
parameters $\mathrm{a,b}$ and $\mathrm{c}$\ are real numbers used as a trick
to recover particular limits. Note in passing that for the case of $\mathsf{n%
}=\left( 1,1\right) $ superalgebra, these extra parameters should be non
zero and can be taken as
\begin{equation}
\mathrm{a=b}=\mathrm{c=1}.
\end{equation}%
Below we want to consider as well particular cases associated with some
special limits which are captured by the singular values of these
parameters. Of particular interest is the set of cases encoded collectively
by the solution of the relation,%
\begin{equation}
\mathrm{abc\ }=\mathrm{0}.
\end{equation}%
Representations involving these cases are related to world line
supersymmetric representations and boundary conformal field theory. Let us
give below some details.

\subsection{Superspace realization}

\qquad Using standard results on supersymmetric field theories, it is not
difficult to see that the superspace representation of the above
superalgebra is given by the following,
\begin{eqnarray}
D_{+} &=&\partial _{+}-\mathrm{a}\theta _{-}\partial _{++}-\mathrm{c}\theta
_{+}\frac{\partial }{\partial \varsigma },  \notag \\
D_{-} &=&\partial _{-}-\mathrm{b}\theta _{+}\partial _{--}-\mathrm{c}\theta
_{-}\frac{\partial }{\partial \varsigma }  \label{3} \\
\partial _{\pm } &=&\frac{\partial }{\partial \theta ^{\mp }},\qquad
\partial _{\pm \pm }=\frac{\partial }{\partial \sigma ^{\mp }},\qquad
\mathcal{Z}=\frac{\partial }{\partial \varsigma }.  \notag
\end{eqnarray}%
These are superspace differential operators acting on superfields
\begin{equation}
\Phi =\Phi \left( \sigma ,\varsigma ,\theta \right)
\end{equation}%
living on the $\left( 3|2\right) $ dimensional real superspace $\mathbb{R}%
^{\left( 3|2\right) }$ with supercoordinates,
\begin{equation}
\mathrm{Z}=\left( \sigma ^{+},\sigma ^{-},\varsigma \text{ }|\text{ }\theta
^{+},\theta ^{-}\right)  \label{y4}
\end{equation}%
The parameters $\sigma ^{+},\sigma ^{-},\varsigma $ are commuting variables (%
$\mathbb{C}$- numbers) and $\theta ^{\pm }\sim \theta _{\mp }$ are Majorana
Grassmann coordinates fulfilling
\begin{eqnarray}
\left\{ \theta _{\pm },\theta _{\mp }\right\} &=&0,\qquad \left\{ \partial
_{\mp },\theta _{\pm }\right\} =1  \notag \\
\lbrack \sigma ^{\pm },\theta _{\pm }] &=&0,\qquad \lbrack \varsigma ,\theta
_{\pm }]=0.  \label{y5}
\end{eqnarray}%
Note that $\varsigma $ is an extra real variable realizing the charge $%
\mathcal{Z}$ as a translation operator in same way as do $\sigma ^{\mp }$
for energy momentum vector $\mathcal{P}_{\pm \pm }$. The latters are
realized as space time translations $\partial _{\pm \pm }$. Typical
hermitian superfields $\Phi _{s}\left( \sigma ,\varsigma ,\theta ^{\pm
}\right) $ have the following $\theta $- expansion,%
\begin{equation}
\Phi _{s}\left( \sigma ,\varsigma ,\theta ^{\pm }\right) =\phi _{s}(\sigma
,\varsigma )+i\theta _{-}\psi _{s+1}(\sigma ,\varsigma )+i\theta _{+}\psi
_{s-1}(\sigma ,\varsigma )+i\theta _{-}\theta _{+}D_{s}(\sigma ,\varsigma ).
\label{y6}
\end{equation}%
For the special case $s=0$, the scalar superfield $\Phi $ contains two real
bosonic fields, $\phi $ and $D$ and two Majorana fermions $\psi _{\pm }$. We
also have the expansions,%
\begin{eqnarray}
D_{+}\Phi &=&i\psi _{+}-\mathrm{a}\theta _{-}\partial _{++}\phi +\theta
_{+}\left( iD-\mathrm{c}\frac{\partial \phi }{\partial \varsigma }\right)
+i\theta _{-}\theta _{+}\left( \mathrm{c}\frac{\partial \psi _{+}}{\partial
\varsigma }-\mathrm{a}\partial _{++}\psi _{-}\right) ,  \notag \\
D_{-}\Phi &=&i\psi _{-}-\mathrm{b}\theta _{+}\partial _{--}\phi -\theta
_{-}\left( iD+\mathrm{c}\frac{\partial \phi }{\partial \varsigma }\right)
-i\theta _{-}\theta _{+}\left( \mathrm{c}\frac{\partial \psi _{-}}{\partial
\varsigma }-\mathrm{b}\partial _{--}\psi _{+}\right) ,  \label{y60}
\end{eqnarray}%
which can be used to build the supersymmetric component field action, see
also eq(\ref{y7}). Note that we have used,%
\begin{equation}
\left( i\theta _{-}\theta _{+}\right) ^{\dagger }=i\theta _{-}\theta _{+}
\end{equation}%
By making appropriate choices of the parameters \textrm{a}, \textrm{b} and
\textrm{c }in eq(\ref{3}), one distinguishes several cases to recover world
line supersymmetry. Besides the case where $D_{-}$ and $D_{+}$\ and $%
\partial _{--}$, $\partial _{++}$, $\frac{\partial }{\partial \varsigma }$
are identified to $D$ and $\frac{d}{dt}$, i.e
\begin{equation}
D^{2}=\frac{d}{dt},\qquad \left[ D,\frac{d}{dt}\right] =0,
\end{equation}%
we have moreover the three following word line like sub-superalgebras of eqs(%
\ref{y1}):\newline
\textbf{(i)} \textbf{Case} $\mathrm{a=c}=0,$ $\mathrm{b}\neq 0$\newline
This particular case corresponds to the so called $\left( 0,1\right) $
heterotic supersymmetry generated by $\left\{ D_{-},\partial _{--}\right\} $%
, i.e%
\begin{eqnarray}
\{D_{+},D_{+}\} &=&0,  \notag \\
\{D_{-},D_{-}\} &=&-2\mathrm{b}\partial _{--} \\
\{D_{+},D_{-}\} &=&0  \notag
\end{eqnarray}%
and can be realized on the (right moving) world line superspace $\mathbb{R}%
^{\left( 1|1\right) }$ parameterized by $\left( \sigma ^{-}|\theta
^{-}\right) $. We have%
\begin{equation}
D_{+}=\partial _{+},\qquad D_{-}=\partial _{-}-\mathrm{b}\theta _{+}\partial
_{--},
\end{equation}%
with $\theta _{+}^{2}=0$. \newline
\textbf{(ii)} \textbf{Case} $\mathrm{b=c}=0,$ $\mathrm{a}\neq 0$\newline
This situation corresponds to $\left( 1,0\right) $ heterotic supersymmetry%
\begin{equation}
\{D_{+},D_{+}\}=-2\mathrm{a}\partial _{++},
\end{equation}%
generated by $D_{+}$ and $\partial _{++}$ and is realized on a (left moving)
world line superspace $\mathbb{R}^{\left( 1|1\right) }$ as well. The
structure of the supersymmetric field theory based on this case is analogous
to the previous one.\newline
\textbf{(iii)} \textbf{Case} $\mathrm{a=b}=0,$ $\mathrm{c}\neq 0$\newline
In this situation, there is no "light cone" energy momentum,%
\begin{equation}
\mathcal{P}_{++}=\mathcal{P}_{--}=0,
\end{equation}%
but we do have a non zero central charge contribution. The corresponding
world line superalgebra
\begin{eqnarray}
\{D_{+},D_{+}\} &=&0,\qquad  \notag \\
\{D_{-},D_{-}\} &=&0 \\
\{D_{\pm },D_{\mp }\} &=&-2\mathrm{c}\mathcal{Z}\text{.}  \notag
\end{eqnarray}%
This special sub-superalgebra of (\ref{y1}) generated by $\left\{ D_{\pm },%
\mathcal{Z}\right\} $ and is realized on a world line superspace $\mathbb{R}%
^{\left( 1|2\right) }$ parameterized by $\left( \varsigma \text{ }|\text{ }%
\theta ^{+},\theta ^{-}\right) $ as follows,
\begin{eqnarray}
D_{+} &=&\partial _{+}-\mathrm{c}\text{ }\theta _{+}\frac{\partial }{%
\partial \varsigma },  \notag \\
D_{-} &=&\partial _{-}-\mathrm{c}\text{ }\theta _{-}\frac{\partial }{%
\partial \varsigma } \\
\mathcal{Z} &=&\frac{\partial }{\partial \varsigma }.  \notag
\end{eqnarray}%
Here there is no dependence in $\sigma ^{\pm }$ variables.

\subsection{Superfield action}

\qquad According to which kind of the above sub-symmetries one is interested
in, we can write down various world line type supersymmetric field actions.
These superfield models should, a priori, be also derived by starting from
the following parent superspace action
\begin{equation}
S\sim \int_{\mathbb{R}^{\left( 3|2\right) }}d^{2}\sigma d\varsigma
d^{2}\theta \left[ \frac{1}{2}D_{-}\Phi D_{+}\Phi +iW(\Phi )\right] .
\label{y7}
\end{equation}%
Then make appropriate reductions as outlined above by keeping track with the
implementation of the constraint eqs to reconstitue missing parts in kinetic
terms of $\left( 1,0\right) $ and $\left( 0,1\right) $ heterotic models. In
eq(\ref{y7}), $W(\Phi )$ is the superpotential; in general given by a
polynom in $\Phi $; but here we shall consider a free field theory and so we
can ignore it.

\qquad Putting the expansion (\ref{y6}) back into this superfield action, we
get after integrating out the Grassmann variables $\theta _{\pm }$ the
following component fields action%
\begin{eqnarray}
S &\sim &\int_{\mathbb{R}^{3}}d^{2}\sigma d\varsigma \left[ \frac{\mathrm{ab}%
}{2}\partial _{--}\phi \partial _{++}\phi +\frac{\mathrm{c}^{2}}{2}\left(
\frac{\partial \phi }{\partial \varsigma }\right) ^{2}-\frac{1}{2}D^{2}%
\right]  \notag \\
&&+\int_{\mathbb{R}^{3}}d^{2}\sigma d\varsigma \left[ \frac{\mathrm{a}}{2}%
\psi _{-}\partial _{++}\psi _{-}-\frac{\mathrm{b}}{2}\psi _{+}\partial
_{--}\psi _{+}\right]  \label{y8} \\
&&+\int_{\mathbb{R}^{3}}d^{2}\sigma d\varsigma \left[ \frac{\mathrm{c}}{2}%
\left( \psi _{+}\frac{\partial }{\partial \varsigma }\psi _{-}-\psi _{-}%
\frac{\partial }{\partial \varsigma }\psi _{+}\right) \right] ,  \notag
\end{eqnarray}%
where we have exhibited the dependence in the \textrm{a}, \textrm{b} and
\textrm{c }moduli. Note that for the particular case where
\begin{equation}
\mathrm{a=b=0,\qquad c=1}
\end{equation}%
the above action reduces, roughly, to the following world line one,%
\begin{eqnarray}
S_{reduced}\left[ \phi ,\psi _{\pm }\right] &\sim &\int_{\mathbb{R}%
}d\varsigma \left[ \frac{\mathrm{1}}{2}\left( \frac{\partial \phi }{\partial
\varsigma }\right) ^{2}-\frac{1}{2}D^{2}\right]  \notag \\
&&+\int_{\mathbb{R}}d\varsigma \left[ \frac{\mathrm{1}}{2}\left( \psi _{+}%
\frac{\partial }{\partial \varsigma }\psi _{-}-\psi _{-}\frac{\partial }{%
\partial \varsigma }\psi _{+}\right) \right] ,  \label{cet}
\end{eqnarray}%
where now $\phi =\phi \left( \varsigma \right) $ and $\psi _{\pm }=\psi
_{\pm }\left( \varsigma \right) $. Note also that as expected, the auxiliary
field D is non dynamical and so can be eliminated through its equation of
motion, an operation once done leads to an on-shell supersymmetric field
action.

Furthermore letting the supercharges act on the superfield $\Phi $, we can
immediately get the infinitesimal supersymmetric transformations,
\begin{eqnarray}
\delta \phi &\sim &i\varepsilon _{-}\psi _{+}+i\varepsilon _{+}\psi _{-}
\notag \\
\delta \psi _{+} &\sim &i\mathrm{a}\varepsilon _{-}\partial _{++}\phi
+\varepsilon _{+}\left( iD+\mathrm{c}\frac{\partial \phi }{\partial
\varsigma }\right)  \notag \\
\delta \psi _{-} &\sim &i\mathrm{b}\varepsilon _{+}\partial _{--}\phi
-\varepsilon _{-}\left( iD-\mathrm{c}\frac{\partial \phi }{\partial
\varsigma }\right) \\
\delta D &\sim &\varepsilon _{-}\left( \mathrm{c}\frac{\partial \psi _{+}}{%
\partial \varsigma }-\mathrm{a}\partial _{++}\psi _{-}\right) -i\varepsilon
_{+}\left( \mathrm{c}\frac{\partial \psi _{-}}{\partial \varsigma }-\mathrm{b%
}\partial _{--}\psi _{+}\right)  \notag
\end{eqnarray}%
With these tools at hand, we turn now to the main objectif of the present
study.

\section{Fermionic like model}

\qquad First we consider the pure bosonic case where the role of the generic
field $\phi $ will be played by the target space time coordinates $x^{m}$
and their conjugate momenta $p_{m}$. Then, we study the derivation of the
pure fermionic analog of the Penrose twistor model. In this case, the role
of world line fermions $\psi _{\pm }$ is played by the four components world
line field $\Upsilon _{\pm }^{m}$.

\subsection{Standard twistor model: Review and comments}

\qquad Roughly, twistor target space formalism allows an alternative
description of the dynamics of a free massless bosonic particle with 4-
energy momentum vector $p_{m}$ satisfying the usual massless condition $\eta
^{mn}p_{m}p_{n}=0$ with $\eta ^{mn}$ the (inverse) metric of the real $%
\mathbb{R}^{\left( \mathrm{d},4-\mathrm{d}\right) }$ space. As $\eta _{mn}$
may have three possible configuration in one to one with the values
\begin{equation}
\mathrm{d}=0,\text{ }1,\text{ }2,
\end{equation}%
we will discuss a special example namely the case of Minkowski space $%
\mathbb{R}^{\left( 1,3\right) }$. But the formalism we will give can be also
done for the two other cases: $\mathbb{R}^{4}$ and $\mathbb{R}^{\left(
\mathrm{2},\mathrm{2}\right) }$. To derive the twistor gauge model from the
usual Minkowski space formulation, we have to proceed in steps. The key idea
can be summarized as follows:

\subsubsection{bi-spinors}

\qquad The first step consists on using $SO(1,3)\simeq SL(2,C)$ isomorphism
to express $4$- vectors of $\mathbb{R}^{\left( 1,3\right) }$ in terms of $%
SL(2,C)$ spinors. Instead of phase space coordinates $x^{m}$ and $p_{m}$,
which can be also written as $\left( \frac{1}{2},\frac{1}{2}\right) $
bi-spinors, twistor formalism uses rather
\begin{equation}
\left( \frac{1}{2},0\right) ,\qquad \text{and}\qquad \left( 0,\frac{1}{2}%
\right) ,
\end{equation}%
spinor fields as fundamental field variables. Before going through technical
details, note first the following useful features: \newline
\textbf{(i)} Given two spinors $\lambda _{1}^{\alpha }$ and $\lambda
_{2}^{\beta }$ , $\alpha ,\beta =1,2$, of positive chirality,%
\begin{equation}
\lambda ^{\alpha }\sim \left( \frac{1}{2},0\right) ,
\end{equation}%
we define the following $SL(2,C)$ invariant
\begin{equation}
\lambda _{1}.\lambda _{2}\text{ }=\text{ }\epsilon _{\alpha \beta }\lambda
_{1}^{\alpha }\lambda _{2}^{\beta },\qquad \epsilon _{\alpha \beta
}=-\epsilon _{\beta \alpha }  \label{a1}
\end{equation}%
Similarly, given two spinors $\overline{\lambda }_{\dot{\alpha}1}$ and $%
\overline{\lambda }_{\dot{\beta}2}$ of negative chirality transforming in $%
\left( 0,\frac{1}{2}\right) $ representation, we can build a second $SL(2,C)$
invariant using dotted variables,
\begin{equation}
\overline{\lambda }_{1}.\overline{\lambda }_{2}=\epsilon ^{\dot{\alpha}\dot{%
\beta}}\overline{\lambda }_{\dot{\alpha}1}\overline{\lambda }_{\dot{\beta}%
2},\qquad \overline{\left( \epsilon ^{\alpha \beta }\right) }=\epsilon _{%
\dot{\beta}\dot{\alpha}}.  \label{a2}
\end{equation}%
For the $SO(3,1)$ case, eqs(\ref{a1}) and (\ref{a2}) are related by usual
complex conjugation ($\overline{\left( \lambda ^{\alpha }\right) }=\overline{%
\lambda }_{\dot{\alpha}}$ and so on); but in the general $SO(\mathrm{p,q})$
case ($\mathrm{p+q=4}$), one has to implement moreover the special
properties of the corresponding spinors.\newline
\textbf{(ii)} Using $SL(2,C)$ spinors, energy momentum vector $p_{m}$ may be
represented line in eq(\ref{d1}). In bi-spinor notations, we have
\begin{equation}
p_{\dot{\beta}}^{\alpha }=\lambda ^{\alpha }\overline{\lambda }_{\dot{\beta}%
}.  \label{ap}
\end{equation}%
Note that the correspondence
\begin{equation*}
p_{m}\qquad \longleftrightarrow \qquad \left( \lambda ,\overline{\lambda }%
\right)
\end{equation*}%
is not a one to one; since the gauge change
\begin{eqnarray}
\lambda ^{\alpha }\qquad &\longrightarrow &\qquad e^{i\theta }\lambda
^{\alpha }  \notag \\
\overline{\lambda }_{\dot{\beta}}\qquad &\longrightarrow &\qquad e^{-i\theta
}\overline{\lambda }_{\dot{\beta}}  \label{tra}
\end{eqnarray}%
with $\theta \in \mathbb{R}$, leaves eq(\ref{ap}) invariant. This symmetry
will play a crucial role in the Penrose twistor model.

\subsubsection{Twistor gauge theory}

\qquad In the second step we introduce Penrose variables $\left( \lambda
^{\alpha },\mathrm{\mu }^{\dot{\alpha}}\right) $ parameterizing the
projective twistor space $PT=CP^{3}$. Their complex conjugates $\left(
\overline{\mathrm{\mu }}_{\alpha },\overline{\lambda }_{\dot{\alpha}}\right)
$ parameterize the anti-holomorphic sector. The coordinates $\left( \lambda
^{\alpha },\mathrm{\mu }^{\dot{\alpha}}\right) $ are commuting phase space
variables taken in the fundamental representations of the $SL\left(
4,C\right) $ twistor group. The isodoublet $\lambda ^{\alpha }$ is as before
while
\begin{equation}
\mathrm{\mu }^{\dot{\alpha}}=\frac{\delta \mathcal{S}_{twistor}}{\delta
\left( \partial _{\tau }\overline{\lambda }_{\dot{\alpha}}\right) }
\end{equation}%
is the conjugate momentum of $\overline{\lambda }_{\dot{\alpha}}$ with $%
\mathcal{S}_{twistor}$ being the twistor field action to be given later on.
Quantum mechanically, the variable $\mathrm{\mu }^{\dot{\beta}}$ gets
promoted to an operator ${\large \mu }^{\dot{\beta}}$ satisfying the
canonical commutation relation,%
\begin{equation}
\left[ \overline{\lambda }_{\dot{\alpha}},{\large \mu }^{\dot{\beta}}\right]
=\delta _{\dot{\alpha}}^{\dot{\beta}}  \label{con}
\end{equation}%
and may be defined as,%
\begin{equation}
{\large \mu }^{\dot{\alpha}}=-\frac{\partial }{\partial \overline{\lambda }_{%
\dot{\alpha}}}.
\end{equation}%
We will turn later to the implication of these quantum equations, for the
moment note that in the classical setting, the geometric points $x_{\beta }^{%
\dot{\alpha}}$ of the real $4D$ Minkowski space-time are associated with the
so-called incidence relation,
\begin{equation}
\mathrm{\mu }^{\dot{\alpha}}=ix_{\beta }^{\dot{\alpha}}\lambda ^{\beta
},\qquad \overline{\mathrm{\mu }}_{\alpha }=i\overline{\lambda }_{\dot{\beta}%
}x_{\alpha }^{\dot{\beta}},\qquad \overline{\left( x_{\beta }^{\dot{\alpha}%
}\right) }=x_{\alpha }^{\dot{\beta}},  \label{inc}
\end{equation}%
describing a line in the spinor space with "slope" $\left( x_{\beta }^{\dot{%
\alpha}}\right) $. These eqs may be also viewed as the solution of the
twistor constraint eq,%
\begin{equation}
\overline{\lambda }_{\dot{\alpha}}\mathrm{\mu }^{\dot{\alpha}}-\lambda
^{\alpha }\overline{\mathrm{\mu }}_{\alpha }=0,
\end{equation}%
describing a helicity zero massless particle. To see this property, it is
interesting to denote collectively the twistor variables as
\begin{equation}
Z^{\mathrm{a}}=\left( \lambda ^{\alpha },\mathrm{\mu }^{\dot{\alpha}}\right)
,  \label{za}
\end{equation}%
with index $\mathrm{a}=\alpha ,$ $\dot{\alpha}$ and projective symmetry as%
\begin{equation}
Z^{\mathrm{a}}\qquad \longrightarrow \qquad Z^{\mathrm{a}\prime }=e^{i\theta
}Z^{\mathrm{a}}  \label{pr}
\end{equation}%
Rising and lowering index $\mathrm{a}$ can be done by help of the metric%
\begin{equation}
\Sigma _{\mathrm{ab}}=\left(
\begin{array}{cc}
0 & \epsilon _{\dot{\alpha}\dot{\beta}} \\
\epsilon _{\alpha \beta } & 0%
\end{array}%
\right) ,\qquad \Sigma _{\mathrm{ab}}\Sigma ^{\mathrm{bc}}=\delta _{\mathrm{a%
}}^{\mathrm{c}}.
\end{equation}%
For instance, we have
\begin{equation}
Z_{\mathrm{a}}=\Sigma _{\mathrm{ab}}Z^{\mathrm{b}}=\left( \mathrm{\mu }_{%
\dot{\alpha}},\lambda _{\alpha }\right) .
\end{equation}%
Note that the twistor variable $Z^{\mathrm{a}}$ and its complex conjugate
\begin{equation}
\left( Z^{\mathrm{a}}\right) ^{\dagger }=\overline{Z}_{\mathrm{a}}=\left( -%
\overline{\mathrm{\mu }}_{\alpha },\overline{\lambda }_{\dot{\alpha}}\right)
\end{equation}%
are respectively in $4$ and $\bar{4}$ representations of the non compact
group $SU(2,2,C)$ $\simeq SO(4,2,\mathbb{R})$. The latter is the conformal
symmetry of the massless (spinless) relativistic particle in $4D$ space-time
and appears as a hidden invariance in the underlying Minkowski space
lagrangian density. Notice moreover that using eqs(\ref{inc}), the following
$SU(2,2)$ invariant%
\begin{equation}
\overline{Z}_{\mathrm{a}}Z^{\mathrm{a}}=\overline{\lambda }_{\dot{\alpha}}%
\mathrm{\mu }^{\dot{\alpha}}-\lambda ^{\alpha }\overline{\mathrm{\mu }}%
_{\alpha }  \label{co}
\end{equation}%
vanishes identically
\begin{equation}
\overline{Z}_{\mathrm{a}}Z^{\mathrm{a}}=0  \label{coo}
\end{equation}%
This constraint eq, which will be derived rigorously later on, can be
checked directly by substituting directly $\mathrm{\mu }^{\dot{\alpha}}$ and
$\overline{\mathrm{\mu }}_{\alpha }$ by their expressions (\ref{inc}). We
have $\overline{\lambda }_{\dot{\alpha}}\mathrm{\mu }^{\dot{\alpha}}=$ $%
i\left( \overline{\lambda }_{\dot{\alpha}}x_{\beta }^{\dot{\alpha}}\lambda
^{\beta }\right) $ which compensates exactly $\overline{\mathrm{\mu }}%
_{\alpha }\lambda ^{\alpha }=i\left( \overline{\lambda }_{\dot{\beta}%
}x_{\alpha }^{\dot{\beta}}\lambda ^{\alpha }\right) $.

\qquad To derive eq(\ref{coo}), we have to identify first the twistor field
action $\mathcal{S}_{twistor}=\mathcal{S}\left[ Z,\overline{Z}\right] $. By
solving $\mathrm{\mu }^{\dot{\alpha}}=\frac{\delta \mathcal{S}_{twistor}}{%
\partial _{\tau }\overline{\lambda }_{\dot{\alpha}}}$ eq(\ref{con}) and its
complex conjugate and using eqs(\ref{za}), it is not difficult to see that
the manifestly $SU(2,2)$ invariant $\mathcal{S}_{twistor}$ reads as follows,%
\begin{equation}
\mathcal{S}_{twistor}\left[ Z,\overline{Z}\right] =-\int d\tau \left[
\overline{Z}_{\mathrm{a}}i\frac{\partial }{\partial \tau }Z^{\mathrm{a}}%
\right] ,  \label{s}
\end{equation}%
To check that this relation is indeed the field action of a massless
particle of Minkowski space, we first replace $\overline{Z}_{\mathrm{a}}$
and $Z^{\mathrm{a}}$ by $\lambda $'s and $\mathrm{\mu }$'s in $\overline{Z}_{%
\mathrm{a}}i\frac{\partial }{\partial \tau }Z^{\mathrm{a}}$. This operation
leads to $\left( \overline{\lambda }_{\dot{\alpha}}i\partial _{\tau }\mathrm{%
\mu }^{\dot{\alpha}}-\overline{\mathrm{\mu }}_{\alpha }i\partial _{\tau
}\lambda ^{\alpha }\right) $. Then use incidence relation (\ref{inc}) to put
the action in the form $\int d\tau \left( \lambda ^{\beta }\overline{\lambda
}_{\dot{\alpha}}\right) \left( \partial _{\tau }x_{\beta }^{\dot{\alpha}%
}\right) $ which is nothing but
\begin{equation}
\mathcal{S}_{Minkowski}=\int d\tau \left( p_{\mu }\partial _{\tau }x^{\mu
}\right) ,  \label{bose}
\end{equation}%
the usual Minkowski space action of a free massless particle.

Moreover using eq(\ref{s}) and requiring gauge invariance under local
projective symmetry (\ref{pr}), the gauge invariant twistor action becomes
\begin{equation}
\mathcal{S}_{twistor}\left[ Z,\overline{Z},V\right] =-\int d\tau \left[
\overline{Z}_{\mathrm{a}}i\left( \partial _{\tau }-iV\right) Z^{\mathrm{a}%
}-2hV\right]
\end{equation}%
where $V$ is a gauge field and $h$ is a Fayet-Iliopoulous (FI) like coupling
constant; it is just the helicity of the spinning particle. By eliminating
the gauge field $V$ through its equation of motion
\begin{equation}
\frac{\delta \mathcal{S}_{twistor}\left[ V\right] }{\delta V}=0,
\end{equation}%
we rediscover the classical constraint eq(\ref{co}) $\sum_{\mathrm{a}}%
\overline{Z}_{\mathrm{a}}Z^{\mathrm{a}}=2h$ which we prefer to rewrite it as
follows,%
\begin{equation}
\sum_{A}\left( Z^{\mathrm{a}}\overline{Z}_{\mathrm{a}}+\overline{Z}_{\mathrm{%
a}}Z^{\mathrm{a}}\right) =4h  \label{cs}
\end{equation}%
and from which one recognizes eq(\ref{co}) once $h$ is set to zero.

\qquad We end this overview and comments on Penrose twistor gauge theory by
saying few words about quantum mechanics. In covariant canonical
quantization of above twistor model, the twistor variable $\overline{Z}_{%
\mathrm{a}}$ (the conjugate momentum of $Z^{\mathrm{a}}$) gets replaced by $-%
\frac{\partial }{\partial Z^{\mathrm{a}}}$. As such the twistor space wave
functions $\Psi (Z)=$ $\left\langle Z\mid \Psi \right\rangle $ are then
holomorphic and homogeneous functions of degree $(-2h-2)$.
\begin{equation}
\Psi (e^{i\theta }Z)=e^{-i\left( 2h+2\right) \theta }\Psi (Z).
\end{equation}%
This homogeneity feature follows from the quantum promotion of the
constraint eq(\ref{cs}) which gets then replaced by,%
\begin{equation}
Z^{\mathrm{a}}\frac{\partial }{\partial Z^{\mathrm{a}}}\Psi (Z)=(-2h-2)\Psi
(Z)  \label{es}
\end{equation}%
We turn now to build our fermionic model.

\subsection{Fermionic construction}

\qquad This pure fermionic construction is important for the derivation of
fermionic twistor like theory to be developed in next section. The field
model we study here deals with the dynamics of world line fermions
transforming as well in the 4- dimension space time and can be viewed as:
\newline
(\textbf{i}) the fermionic analog of eqs(\ref{seb},\ref{bose}) dual to
twistor construction \`{a} la Penrose.\newline
(\textbf{ii}) the fermionic sector of a supersymmetric gauge theory along
the lines outlined in section 2.\newline
Instead of the world line field bosons $x_{\beta }^{\dot{\alpha}}\left( \tau
\right) $ and $p_{\dot{\alpha}}^{\beta }\left( \tau \right) $ considered
above, our pure fermionic model involves the following ingredients:\newline
(\textbf{1}) \textbf{Degrees of freedom}:\qquad The model involves the world
line fermions
\begin{equation}
\Upsilon _{-\beta }^{\dot{\alpha}}\left( \tau \right) ,\qquad \Upsilon
_{+\beta }^{\dot{\alpha}}\left( \tau \right) ,\qquad \widetilde{\Upsilon }%
_{-\beta }^{\dot{\alpha}}\left( \tau \right) ,\qquad \widetilde{\Upsilon }%
_{+\beta }^{\dot{\alpha}}\left( \tau \right)
\end{equation}%
(see eqs(\ref{gp}) to fix the ideas). These fields carries two kinds of
indices: (\textbf{a}) An index ($\pm $) exhibiting the statistics of $%
\Upsilon _{\pm \beta }^{\dot{\alpha}}$ and $\widetilde{\Upsilon }_{\pm \beta
}^{\dot{\alpha}}$; they are world line fermions. (\textbf{b}) dotted and
undotted indices ($\alpha ,\dot{\alpha}=1,2$) showing that the fields
transform as 4- vectors (bi-spinors) in one of the three possible $\mathbb{R}%
^{\left( d,4-d\right) }$ spaces,%
\begin{equation}
\Upsilon _{\pm }^{m}=\left( \sigma ^{m}\right) _{\dot{\beta}}^{\alpha
}\Upsilon _{\pm \alpha }^{\dot{\beta}},\qquad \widetilde{\Upsilon }_{\pm
}^{m}=\left( \sigma ^{m}\right) _{\dot{\beta}}^{\alpha }\widetilde{\Upsilon }%
_{\pm \alpha }^{\dot{\beta}}.  \label{x1}
\end{equation}%
Though stringy applications suggest to consider $\mathbb{R}^{\left(
1,3\right) }$, it is also interesting to consider also the $\mathbb{R}%
^{\left( 2,2\right) }$ and $\mathbb{R}^{4}$ cases. Below, we shall mainly
focus our attention here on the case $\mathbb{R}^{\left( 2,2\right) }$
because of specific properties; in particular because of the two following:%
\newline
(\textbf{i}) In $\mathbb{R}^{\left( 1,3\right) }$ and $\mathbb{R}^{4}$ with
rotation groups $SO\left( 1,3,\mathbb{R}\right) \simeq SL\left( 2,\mathbb{C}%
\right) $ and $SO\left( 4,\mathbb{R}\right) \simeq $ $SU\left( 2\right)
\times $ $SU\left( 2\right) $ respectively, $\left( \frac{1}{2},0\right) $
and $\left( 0,\frac{1}{2}\right) $\ spinors are complex,%
\begin{equation}
\widetilde{\Upsilon }_{\pm }^{m}=\overline{\left( \Upsilon _{\pm
}^{m}\right) }.  \label{bar}
\end{equation}%
and the underlying twistor space $PT$ splits into holomorphic and
antiholomorphic sectors interchanged under complex conjugation,
\begin{equation}
\left( \frac{1}{2},0\right) \qquad \longleftrightarrow \qquad \left( 0,\frac{%
1}{2}\right)
\end{equation}%
This complex structure is rich from algebraic geometry view and stringy
applications, in particular in topological string on conifold.\newline
(\textbf{ii}) In $\mathbb{R}^{\left( 2,2\right) }$ the rotation group is
given by $SO\left( 2,2,\mathbb{R}\right) \simeq SL\left( 2,\mathbb{R}\right)
\times SL\left( 2,\mathbb{R}\right) $ and basic spinors are real. In this
case, it is possible to drop out half of the degrees of freedom by keeping
for instance just
\begin{equation}
\Upsilon _{\pm \beta }^{\dot{\alpha}}\sim \Upsilon _{\pm }^{m},
\end{equation}%
in eqs(\ref{x1}).

Note also that since $\left( \Upsilon _{\pm }^{m}\right) =\left( \Upsilon
_{\pm }^{0},\Upsilon _{\pm }^{1},\Upsilon _{\pm }^{2},\Upsilon _{\pm
}^{4},\right) $ are fermions, the square of each component $\Upsilon _{\pm
}^{m}$ vanishes individually,%
\begin{equation}
\left( \Upsilon _{\pm }^{m}\right) ^{2}=0,\qquad m=0,1,2,3.  \label{x2}
\end{equation}%
Moreover since the metric $\eta _{mn}$ is diagonal we also have%
\begin{equation}
\eta _{mn}\Upsilon _{\pm }^{m}\Upsilon _{\pm }^{n}\sim \epsilon ^{\alpha
\beta }\epsilon _{\dot{\alpha}\dot{\beta}}\Upsilon _{-\alpha }^{\dot{\alpha}%
}\Upsilon _{+\beta }^{\dot{\beta}}=0.  \label{x3}
\end{equation}%
Similar relations may be written down for $\widetilde{\Upsilon }_{\pm }^{m}$.%
\newline
(\textbf{2}) \textbf{Field action}:\qquad The pure fermionic field action $%
\mathcal{S}_{F}$ describing the dynamics of the free fields $\Upsilon _{\pm
\alpha }^{\dot{\alpha}}$ and $\widetilde{\Upsilon }_{-\alpha }^{\dot{\alpha}%
} $ is given by, see also eq(\ref{cet}),%
\begin{eqnarray}
\mathcal{S}_{F}\left[ \Upsilon ,\widetilde{\Upsilon }\right] &=&\int_{%
\mathbb{R}}d\tau \left[ \frac{\mathrm{1}}{2}\left( \Upsilon _{-\alpha }^{%
\dot{\alpha}}\frac{\partial }{\partial \tau }\Upsilon _{+\beta }^{\dot{\beta}%
}-\Upsilon _{+\beta }^{\dot{\beta}}\frac{\partial }{\partial \tau }\Upsilon
_{-\alpha }^{\dot{\alpha}}\right) \epsilon ^{\alpha \beta }\epsilon _{\dot{%
\alpha}\dot{\beta}}\right]  \notag \\
&&+\int_{\mathbb{R}}d\tau \left[ \frac{\mathrm{1}}{2}\left( \widetilde{%
\Upsilon }_{-\alpha }^{\dot{\alpha}}\frac{\partial }{\partial \tau }%
\widetilde{\Upsilon }_{+\beta }^{\dot{\beta}}-\widetilde{\Upsilon }_{+\beta
}^{\dot{\beta}}\frac{\partial }{\partial \tau }\widetilde{\Upsilon }%
_{-\alpha }^{\dot{\alpha}}\right) \epsilon ^{\alpha \beta }\epsilon _{\dot{%
\alpha}\dot{\beta}}\right] .  \label{x4}
\end{eqnarray}%
Note in passing that, under some hypothesis on chirality, the second line
may dropped out. Note also that comparing $\mathcal{S}_{F}$ , which can be
also written as
\begin{eqnarray}
\mathcal{S}_{F}\left[ \Upsilon _{\pm },\partial _{\tau }\Upsilon _{\pm }%
\right] &\sim &\int_{\mathbb{R}}d\tau \left[ \eta _{mn}\left( \Upsilon
_{-}^{m}\frac{\partial }{\partial \tau }\Upsilon _{+}^{n}\right) \right]
\notag \\
&&+\int_{\mathbb{R}}d\tau \left[ \eta _{mn}\left( \widetilde{\Upsilon }%
_{-}^{m}\frac{\partial }{\partial \tau }\widetilde{\Upsilon }_{+}^{n}\right) %
\right] ,  \label{x5}
\end{eqnarray}%
with its Bose analog eq(\ref{bose}), one learns the naive correspondence,%
\begin{eqnarray}
x^{n}\qquad &\longrightarrow &\qquad \Upsilon _{+}^{n},\text{ \ }\widetilde{%
\Upsilon }_{+}^{n}  \notag \\
p_{m}\qquad &\longrightarrow &\qquad \Upsilon _{-m},\text{ \ }\widetilde{%
\Upsilon }_{-m}.  \label{x6}
\end{eqnarray}%
(\textbf{3}) The conjugate momentum variables ${\Large \pi }_{-\dot{\alpha}%
}^{\beta }$ and $\widetilde{{\Large \pi }}_{-\dot{\alpha}}^{\beta }$,%
\begin{equation}
{\Large \pi }_{-\dot{\alpha}}^{\beta }\left( \tau \right) =\frac{\delta
\mathcal{S}_{F}}{\delta \left( \partial \Upsilon _{+\beta }^{\dot{\alpha}%
}\right) },\qquad \widetilde{{\Large \pi }}_{-\dot{\alpha}}^{\beta }=\frac{%
\delta \mathcal{S}_{F}}{\delta \left( \partial \widetilde{\Upsilon }_{+\beta
}^{\dot{\alpha}}\right) }.  \label{b1}
\end{equation}%
By using the above expression of $\mathcal{S}_{F}$, we get%
\begin{equation}
{\Large \pi }_{-\dot{\alpha}}^{\alpha }=-\epsilon ^{\alpha \beta }\epsilon _{%
\dot{\alpha}\dot{\beta}}\Upsilon _{-\beta }^{\dot{\beta}},\qquad \widetilde{%
{\Large \pi }}_{-\dot{\alpha}}^{\alpha }=-\epsilon ^{\alpha \beta }\epsilon
_{\dot{\alpha}\dot{\beta}}\widetilde{\Upsilon }_{-\beta }^{\dot{\beta}}.
\end{equation}%
This relation shows that the fermionic fields $\Upsilon _{+\beta }^{\dot{%
\beta}}$, $\widetilde{\Upsilon }_{+\beta }^{\dot{\beta}}$\ and $\Upsilon _{-%
\dot{\beta}}^{\beta }$, $\widetilde{\Upsilon }_{-\dot{\beta}}^{\beta }$\ are
conjugate to each other. So to avoid confusions, it is interesting to
re-parameterize the above fermionic field action as follows:%
\begin{equation}
\mathcal{S}_{F}\left[ \Upsilon ,{\Large \pi ,}\widetilde{\Upsilon },%
\widetilde{{\Large \pi }}\right] =\int_{\mathbb{R}}d\tau \mathrm{Tr}\left(
{\Large \pi }_{-\dot{\alpha}}^{\alpha }\frac{\partial }{\partial \tau }%
\Upsilon _{+\alpha }^{\dot{\alpha}}+\widetilde{{\Large \pi }}_{-\dot{\alpha}%
}^{\alpha }\frac{\partial }{\partial \tau }\widetilde{\Upsilon }_{+\alpha }^{%
\dot{\alpha}}\right)  \label{x8}
\end{equation}%
Note also because of the fermionic nature of the ${\Large \pi }_{-\dot{\alpha%
}}^{\alpha }$'s and like for eq(\ref{x3}), we have here also the nilpotency
property
\begin{equation}
\mathrm{Tr}\left( {\Large \pi }_{-}^{2}\right) ={\Large \pi }_{-\dot{\alpha}%
}^{\alpha }{\Large \pi }_{-\alpha }^{\dot{\alpha}}=0,\qquad \mathrm{Tr}%
\left( \widetilde{{\Large \pi }}_{-}^{2}\right) =\widetilde{{\Large \pi }}_{-%
\dot{\alpha}}^{\alpha }\widetilde{{\Large \pi }}_{-\alpha }^{\dot{\alpha}}=0.
\label{x9}
\end{equation}%
Since the pairs $\left( \Upsilon _{+\beta }^{\dot{\beta}},{\Large \pi }_{-%
\dot{\alpha}}^{\alpha }\right) $\ and $\left( \widetilde{\Upsilon }_{+\dot{%
\beta}}^{\beta },\widetilde{{\Large \pi }}_{-\dot{\alpha}}^{\alpha }\right) $
play quite similar role, we shall fix our attention on one of them, say $%
\left( \Upsilon _{+\beta }^{\dot{\beta}},{\Large \pi }_{-\dot{\alpha}%
}^{\alpha }\right) $ with field action
\begin{equation}
\mathcal{S}_{F}\left[ \Upsilon ,{\Large \pi }\right] =\int_{\mathbb{R}}d\tau
\mathrm{Tr}\left( {\Large \pi }_{-\dot{\alpha}}^{\alpha }\frac{\partial }{%
\partial \tau }\Upsilon _{+\alpha }^{\dot{\alpha}}\right) ,  \label{x80}
\end{equation}%
then give the results for $\left( \widetilde{\Upsilon }_{+\dot{\beta}%
}^{\beta },\widetilde{{\Large \pi }}_{-\dot{\alpha}}^{\alpha }\right) $
whenever needed.

\subsubsection{Basic symmetries}

\qquad In the case of $\mathbb{R}^{\left( 2,2\right) }$ geometry, the field
action $\mathcal{S}_{F}$ has the following remarkable and manifest
symmetries:\newline
(\textbf{1}) $SO\left( 2,2,\mathbb{R}\right) \sim SL\left( 2,\mathbb{R}%
\right) \times SL\left( 2,\mathbb{R}\right) $ global invariance. \newline
(\textbf{2}) Invariance under the real scaling transformation
\begin{eqnarray}
\Upsilon _{+\alpha }^{\dot{\alpha}}\qquad &\longrightarrow &\qquad
e^{+\upsilon }\Upsilon _{+\alpha }^{\dot{\alpha}}  \notag \\
{\Large \pi }_{-\dot{\alpha}}^{\alpha }\qquad &\longrightarrow &\qquad
e^{-\upsilon }{\Large \pi }_{-\dot{\alpha}}^{\alpha }.  \label{1}
\end{eqnarray}%
with a global real parameter $\upsilon \in \mathbb{R}$. It captures the spin
of the world line fields and has also a world sheet interpretation in string
theory.\newline
Moreover promoting this symmetry to a local one, that is $\upsilon =\upsilon
\left( \tau \right) $ or equivalently,
\begin{equation}
\frac{\partial \upsilon }{\partial \tau }\neq 0,
\end{equation}
then gauge invariance requires that the above field action (\ref{x80})
should be extended as,%
\begin{equation}
\mathcal{S}_{FG}\left[ \Upsilon _{+},{\Large \pi }_{-},\mathcal{A}\right]
\text{ }\sim \text{ }\int d\tau \mathrm{Tr}\left[ {\Large \pi }_{-\dot{\alpha%
}}^{\alpha }\left( \frac{\partial }{\partial \tau }-\mathcal{A}_{\tau
}T_{0}\right) \Upsilon _{+\alpha }^{\dot{\alpha}}\right]  \label{b6}
\end{equation}%
where $\mathcal{A}=\mathcal{A}\left( \tau \right) $ is the gauge field
associated with the above real scaling invariance. In this relation, $T_{0}$
is the generator of the scaling symmetry (\ref{1}) and acts on the fields $%
\Upsilon _{+}$ and ${\Large \pi }_{-}$ like:%
\begin{equation}
\left[ T_{0},\Upsilon _{+\alpha }^{\dot{\alpha}}\right] =+\frac{1}{2}%
\Upsilon _{+\alpha }^{\dot{\alpha}},\qquad \left[ T_{0},{\Large \pi }_{-\dot{%
\alpha}}^{\alpha }\right] =-\frac{1}{2}{\Large \pi }_{-\dot{\alpha}}^{\alpha
}.
\end{equation}%
The meaning of this gauge invariance can be made transparent by noting that
the gauge field $\mathcal{A}_{\tau }$ do not propagate and so it is an
auxiliary field. Eliminating it through its equation of motion
\begin{equation}
\frac{\delta \mathcal{S}_{FG}}{\delta \mathcal{A}_{\tau }}=0,
\end{equation}%
we get the following constraint eq,%
\begin{equation}
\mathrm{Tr}\left( {\Large \pi }_{-\dot{\alpha}}^{\alpha }\Upsilon _{+\alpha
}^{\dot{\alpha}}\right) =0.  \label{b7}
\end{equation}%
But this equation is a familiar QFT relation; it captures the antisymmetry
property of Fermi like fields as it can be seen by using the cyclic property
of trace, $\mathrm{Tr}\left( {\Large \pi }_{-\dot{\alpha}}^{\alpha }\Upsilon
_{+\alpha }^{\dot{\alpha}}+\Upsilon _{+\alpha }^{\dot{\alpha}}{\Large \pi }%
_{-\dot{\alpha}}^{\alpha }\right) =0$.

\begin{theorem}
\ \newline
In pure fermionic phase space with canonical coordinate fields $\Upsilon
_{+\alpha }^{\dot{\alpha}}$ ($\widetilde{\Upsilon }_{+\alpha }^{\dot{\alpha}%
} $) and conjugate momenta ${\Large \pi }_{-\dot{\alpha}}^{\alpha }$ ($%
\widetilde{{\Large \pi }}_{-\dot{\alpha}}^{\alpha }$), the free field action
$\mathcal{S}_{FG}$ ($\widetilde{\mathcal{S}}_{FG}$) has a local $SP\left( 2,%
\mathbb{R}\right) $ gauge invariance capturing the three Fermi field
antisymmetry properties%
\begin{eqnarray}
\mathrm{Tr}\left( {\Large \pi }_{-\dot{\alpha}}^{\alpha }\Upsilon _{+\alpha
}^{\dot{\alpha}}\right) &=&0  \notag \\
\mathrm{Tr}\left( {\Large \pi }_{-\dot{\alpha}}^{\alpha }{\Large \pi }%
_{-\alpha }^{\dot{\alpha}}\right) &=&0  \label{3cst} \\
\mathrm{Tr}\left( \Upsilon _{+\dot{\alpha}}^{\alpha }\Upsilon _{+\alpha }^{%
\dot{\alpha}}\right) &=&0.  \notag
\end{eqnarray}%
These are just the constraint relations derived above; see eqs(\ref{x3},\ref%
{x9},\ref{b7}). Similar relations are also valid for $\widetilde{\Upsilon }%
_{+\dot{\beta}}^{\beta }$ and $\widetilde{{\Large \pi }}_{-\dot{\alpha}%
}^{\alpha }$.
\end{theorem}

\subsubsection{More on $SP\left( 2,\mathbb{R}\right) $ gauge invariance}

\qquad Before giving the $SP\left( 2,\mathbb{R}\right) $ gauge invariant
field action, we begin by recalling some useful tools on the real group $%
SP\left( 2,\mathbb{R}\right) $ and some aspects of its representations. Its
Lie algebra $sp\left( 2,\mathbb{R}\right) $ is three dimensional with
generators $T_{--}$,$T_{0}$ and $T_{++}$ satisfying the commutation
relations,%
\begin{eqnarray}
\left[ T_{--},T_{++}\right] &=&2T_{0},  \notag \\
\left[ T_{0},T_{++}\right] &=&T_{++},\qquad \\
\left[ T_{0},T_{--}\right] &=&-T_{--},\qquad  \notag
\end{eqnarray}%
where $\Delta =2T_{0}$ is the operator counting the charges of $SP\left( 2,%
\mathbb{R}\right) $ field representations. In particular we have,
\begin{equation}
\left[ \Delta ,T_{\pm \pm }\right] =\pm 2T_{\pm \pm },\qquad \left[ \Delta
,\Upsilon _{+\alpha }^{\dot{\alpha}}\right] =+\Upsilon _{+\alpha }^{\dot{%
\alpha}},\qquad \left[ \Delta ,{\Large \pi }_{-\dot{\alpha}}^{\alpha }\right]
=-{\Large \pi }_{-\dot{\alpha}}^{\alpha }.
\end{equation}%
There are different ways to realize this real symmetry; one of these ways is
given by the following left moving wold sheet differential operators,%
\begin{equation}
T_{--}=\frac{\partial }{\partial \sigma ^{+}},\qquad T_{0}=\sigma ^{+}\frac{%
\partial }{\partial \sigma ^{+}},\qquad T_{++}=\sigma ^{+2}\frac{\partial }{%
\partial \sigma ^{+}}.
\end{equation}%
from which one can read immediately their scaling behaviours under global
the transformations $\sigma ^{+}\longrightarrow e^{2\upsilon }\sigma ^{+}$,%
\begin{equation}
T_{--}^{\prime }=e^{-2\upsilon }T_{--},\qquad T_{0}^{\prime }=T_{0},\qquad
T_{++}^{\prime }=e^{2\upsilon }T_{++}.
\end{equation}%
The $T_{0,\pm \pm }$'s can be also realized as $2\times 2$ matrices when
acting on two component vectors This is the case which interests us here.
The 2-vector is given by ($\Upsilon _{+},{\Large \pi }_{-}$) and the $%
T_{0,\pm \pm }$ matrices act as,%
\begin{eqnarray}
\left[ T_{++},\Upsilon _{+}\right] &=&\text{ \ }0,\qquad \text{\ \ \ \ \ \ \
\ }\left[ T_{--},{\Large \pi }_{-}\right] =\text{ \ }0  \notag \\
\left[ T_{--},\Upsilon _{+}\right] &\sim &\text{ \ \ }{\Large \pi }%
_{-},\qquad \ \ \ \ \ \left[ T_{++},{\Large \pi }_{-}\right] \sim \text{ \ \
}\Upsilon _{+}  \label{12} \\
\left[ T_{0},\Upsilon _{+}\right] &=&+\frac{1}{2}\Upsilon _{+},\qquad \text{%
\ \ \ }\left[ T_{0},{\Large \pi }_{-}\right] =-\frac{1}{2}{\Large \pi }_{-}
\notag
\end{eqnarray}%
We come now to derive the general expression of the classical field action $%
\mathcal{S}_{FG}\left[ \Upsilon _{+},{\Large \pi }_{-},\mathcal{B}\right] $
extending eq(\ref{b6}) and recovering the constraint eqs(\ref{3cst}). This
action has an $SP\left( 2,\mathbb{R}\right) $ gauge invariance and reads as,
\begin{equation}
\mathcal{S}_{FG}\left[ \Upsilon _{+},{\Large \pi }_{-},\mathcal{B}\right]
\sim \int d\tau \left( \mathrm{Tr}\left[ {\Large \pi }_{-\dot{\alpha}%
}^{\alpha }\left( \frac{\partial }{\partial \tau }-\mathcal{B}_{\tau
}\right) \Upsilon _{+\alpha }^{\dot{\alpha}}\right] \right)  \label{ssg}
\end{equation}%
where now the gauge field $\mathcal{B}_{\tau }$ is valued in the $sp\left( 2,%
\mathbb{R}\right) $\ algebra, i.e,
\begin{equation}
\mathcal{B}_{\tau }=\mathcal{B}_{\tau ,0}T_{0}+\mathcal{B}_{\tau ,--}T_{++}+%
\mathcal{B}_{\tau ,++}T_{--}.
\end{equation}%
Eq(\ref{ssg}) can be put into a more condensed form by using the $SP\left( 2,%
\mathbb{R}\right) $ spinor $\Gamma ^{i}=\epsilon ^{ij}\Gamma _{j}$ with $%
i=+,-$, $\epsilon ^{+-}=1$; that is%
\begin{equation}
\Gamma _{i}=\left( {\Large \pi }_{-},-\Upsilon _{+}\right) ,\qquad \Gamma
^{i}=\left(
\begin{array}{c}
\Upsilon _{+} \\
{\Large \pi }_{-}%
\end{array}%
\right) ,
\end{equation}%
We have%
\begin{equation}
\mathcal{S}_{FG}\left[ \Gamma ,\mathcal{B}\right] \sim \int d\tau \left(
\frac{1}{2}\mathrm{Tr}\left[ \Gamma _{\dot{\alpha}}^{i\alpha }\left(
\epsilon _{ij}\frac{\partial }{\partial \tau }-\left( \mathcal{B}_{\tau
}\right) _{\left( ij\right) }\right) \Gamma _{\alpha }^{j\dot{\alpha}}\right]
\right)
\end{equation}%
From which one can re-express the constraint eqs(\ref{3cst}) using the world
line $\Gamma $ fields. Similar results are valid for the twild sector $%
\widetilde{\Gamma }$ and $\widetilde{\mathcal{B}}$.

\section{Twistor like dual}

\qquad To build the twistor representation dual to the above the pure
fermionic model, we proceed as in bosonic case. Below, we summarize the
steps of the procedure:

\subsection{Solve the constraint eqs}

\qquad First solve the constraint eqs(\ref{3cst}) by using fermionic twistor
like variables. Following the idea outlined in introduction and using
preliminary results of section 2; then mimicking the usual bosonic twistor
analysis, it is not difficult to see that the solution of the constraint eqs
is given by,%
\begin{eqnarray}
\Upsilon _{+\alpha }^{\dot{\alpha}} &=&\lambda _{\alpha }\zeta _{+}^{\dot{%
\alpha}},\qquad \widetilde{\Upsilon }_{+\alpha }^{\dot{\alpha}}=\zeta
_{+\alpha }\lambda ^{\dot{\alpha}}  \notag \\
{\Large \pi }_{-\dot{\alpha}}^{\alpha } &=&\lambda ^{\alpha }\zeta _{-\dot{%
\alpha}},\qquad \widetilde{{\Large \pi }}_{-\alpha }^{\dot{\alpha}}=\zeta
_{-\alpha }\lambda ^{\dot{\alpha}},  \label{s1}
\end{eqnarray}%
where, for simplicity, we have set dropped out the hats on the fields; $%
\widetilde{\zeta }_{\pm }^{\dot{\alpha}}\equiv \zeta _{\pm }^{\dot{\alpha}}$
and $\widetilde{\lambda }^{\dot{\alpha}}\equiv \lambda ^{\dot{\alpha}}$. In
this solution, the $\lambda _{\alpha }$'s ($\lambda ^{\dot{\alpha}}$'s) are
commuting world line bosons with the usual property $\left( \lambda _{\alpha
}\lambda ^{\alpha }\right) =0$; the same $\lambda _{\alpha }$'s as in
section 3. The $\zeta _{\pm }^{\dot{\alpha}}$'s ($\zeta _{\pm \alpha }$'s)
are anticommuting world line fermions capturing the Fermi statistics of $%
\Upsilon _{+\alpha }^{\dot{\alpha}}$ and ${\Large \pi }_{-\dot{\alpha}%
}^{\alpha }$ ($\widetilde{\Upsilon }_{+\alpha }^{\dot{\alpha}}$ and $%
\widetilde{{\Large \pi }}_{-\dot{\alpha}}^{\alpha }$). By substituting the
above solution in the anticommutator,
\begin{equation}
\left\{ \Upsilon _{+\alpha }^{\dot{\alpha}},{\Large \pi }_{-\dot{\beta}%
}^{\beta }\right\} =0,
\end{equation}%
we get in a first stage $\left\{ \lambda _{\alpha }\zeta _{+}^{\dot{\alpha}%
},\lambda ^{\beta }\zeta _{-\dot{\beta}}\right\} =\lambda _{\alpha }\lambda
^{\beta }\left\{ \zeta _{+}^{\dot{\alpha}},\zeta _{-\dot{\beta}}\right\} $
and so the following identities,%
\begin{equation}
\left\{ \zeta _{-\dot{\alpha}},\zeta _{+}^{\dot{\beta}}\right\} =0,\qquad %
\left[ \lambda _{\alpha },\lambda ^{\beta }\right] =0,\qquad \left[ \lambda
_{\alpha },\zeta _{\pm }^{\dot{\beta}}\right] =0,
\end{equation}%
showing that $\Upsilon _{+\alpha }^{\dot{\alpha}}$ is built out of the
product of boson $\lambda _{\alpha }$ with positive $SO\left( 2,2,R\right) $
chirality and a world line fermion with negative $SO\left( 2,2,R\right) $\
chirality. The opposite combination corresponds exactly to ${\Large \pi }_{-%
\dot{\alpha}}^{\alpha }$. \newline
Moreover, notice that the solution (\ref{s1}) is not unique since arbitrary
changes of the $SO\left( 2,2,R\right) $ spinors of the form%
\begin{eqnarray}
\lambda ^{\alpha }\qquad &\rightarrow &\qquad \lambda ^{\alpha \prime
}=e^{\varphi }\lambda ^{\alpha },\qquad  \notag \\
\zeta _{\pm }^{\dot{\alpha}}\qquad &\rightarrow &\qquad \zeta _{\pm }^{\dot{%
\alpha}\prime }=e^{-\varphi }\zeta _{\pm }^{\dot{\alpha}},  \label{p1}
\end{eqnarray}%
with $\varphi =\varphi \left( \tau \right) $ is an arbitrary real number; $%
\varphi \in \mathbb{R}$, leaves eqs(\ref{s1}) invariant. This transformation
corresponds to the usual complex projective symmetry of bosonic twistor
space. Notice also that under the scaling transformations $\Upsilon
_{+\alpha }^{\dot{\alpha}}\rightarrow e^{\upsilon }\Upsilon _{+\alpha }^{%
\dot{\alpha}}$ and ${\Large \pi }_{-\dot{\alpha}}^{\alpha }\rightarrow
e^{-\upsilon }{\Large \pi }_{-\dot{\alpha}}^{\alpha }$ eqs(\ref{s1}), we
have moreover,%
\begin{equation}
\lambda ^{\alpha }\qquad \rightarrow \qquad \lambda ^{\alpha },\qquad \zeta
_{\pm }^{\dot{\alpha}}\qquad \rightarrow \qquad e^{\pm \upsilon }\zeta _{\pm
}^{\dot{\alpha}},  \label{sc}
\end{equation}%
and should not be confused with eq(\ref{p1}). As this symmetry captures just
the Fermi statistics of the $\zeta _{\pm }$'s, we shall forget it and then
focus our attention in what follows on the real projective invariance (\ref%
{p1}). This symmetry plays a crucial role in the derivation fermionic
twistor like model and its quantization.

\subsection{Twistor field action $\mathcal{S}_{FT}$}

\qquad To build the fermionic twistor like field action $\mathcal{S}_{FT}$
dual to eq(\ref{x4}) and derive the underlying gauge constraint eqs, we
start by introducing the projective twistor variables
\begin{equation}
z^{a},\text{ }\mathrm{w}_{a},
\end{equation}%
for the fermionic twistor like theory. Next we derive the field action $%
\mathcal{S}_{FT}=\mathcal{S}_{FT}\left[ z^{a},\mathrm{w}_{a}\right] $ of the
classical model and consider then its quantization.

\qquad Roughly, twistor variables are twistor phase space coordinates, given
in the case of a massless particle moving in $\mathbb{R}^{\left( 1,3\right)
} $, by%
\begin{equation}
\lambda ^{\alpha },\text{ }\mathrm{\mu }^{\dot{\alpha}};\text{ }\overline{%
\mathrm{\mu }}_{\alpha },\text{ }\overline{\lambda }_{\dot{\alpha}},
\end{equation}%
where $\mathrm{\mu }^{\dot{\alpha}}=\frac{\delta S_{BT}}{\delta \partial
_{\tau }\overline{\lambda }_{\dot{\alpha}}}$ and $\overline{\mathrm{\mu }}%
_{\alpha }=\frac{\delta S_{BT}}{\delta \partial _{\tau }\lambda ^{\alpha }}$%
. It happens that these twistor variables split into complex homogeneous
holomorphic variables
\begin{equation}
z^{a}=\left( \lambda ^{\alpha },\text{ }\mathrm{\mu }^{\dot{\alpha}}\right)
\text{ \ }\in \text{ \ }CP^{3},
\end{equation}%
and their complex antiholomorphic ones $\overline{z}_{a}=\left( \overline{%
\lambda }_{\dot{\alpha}},\text{ }\overline{\mathrm{\mu }}_{\alpha }\right) $%
. In the case of a massless particle moving in the $\mathbb{R}^{\left(
2,2\right) }$ geometry, the analysis is a little bit different for the two
following:\newline
(\textbf{i}) Spinors of $SO\left( 2,2,\mathbb{R}\right) $ target space group
are real; so the corresponding twistor space is real and apparently there is
no holomorphic and anti-holomorphic like splitting as in the complex case.
However, we do still have the two sectors given by $\left( \Upsilon
_{+\alpha }^{\dot{\alpha}},{\Large \pi }_{-\dot{\alpha}}^{\alpha }\right) $
and $\left( \widetilde{\Upsilon }_{+\alpha }^{\dot{\alpha}},\widetilde{%
{\Large \pi }}_{-\dot{\alpha}}^{\alpha }\right) $, but stable under complex
conjugation. More precisely, we have for the $\left( \Upsilon _{+\alpha }^{%
\dot{\alpha}},{\Large \pi }_{-\dot{\alpha}}^{\alpha }\right) $ sector the
twistor like variables,
\begin{equation}
\left( \lambda ^{1},\text{ }\lambda ^{2},\text{ }\mathrm{\xi }_{-\dot{1}},%
\text{ }\mathrm{\xi }_{-\dot{2}},\text{ }\mathrm{\nu }_{1},\text{ }\mathrm{%
\nu }_{2},\text{ }\zeta _{+}^{\dot{1}},\text{ }\zeta _{+}^{\dot{2}}\right)
\text{ \ }\in \text{ \ }WRP_{\left( +1,+1,+1,+1,-1,-1,-1,-1\right) }^{\left(
3|4\right) },  \label{sec1}
\end{equation}%
where $\left( +1,+1,+1,+1,-1,-1,-1,-1\right) $ stand for the weight vector
of the scaling transformation of the super multiplet $\left( \lambda
^{\alpha },\text{ }\mathrm{\xi }_{-\dot{\alpha}},\text{ }\mathrm{\nu }%
_{\alpha },\text{ }\zeta _{+}^{\dot{\alpha}}\right) $. The space $%
WRP_{\left( ++++----\right) }^{\left( 3|4\right) }$ stands for the weighted
real projective supermanifold. For $\left( \widetilde{\Upsilon }_{+\alpha }^{%
\dot{\alpha}},\widetilde{{\Large \pi }}_{-\dot{\alpha}}^{\alpha }\right) $
sector, we have the following projective supercoordinates,
\begin{equation}
\left( \zeta _{+}^{\alpha },\text{ }\mathrm{\nu }^{\dot{\alpha}},\text{ }%
\mathrm{\xi }_{-\alpha },\text{ }\lambda _{\dot{\alpha}}\right) .
\end{equation}%
In first sector bosons $\lambda ^{\alpha }$ and $\mathrm{\nu }_{\alpha }$
are in $\left( \frac{1}{2},0\right) $ representation of $SO\left( 2,2,%
\mathbb{R}\right) $ while in the second, $\lambda _{\dot{\alpha}}$ and $%
\mathrm{\nu }^{\dot{\alpha}}$ are in $\left( 0,\frac{1}{2}\right) $.
Fermions have the opposite chirality.\newline
\textbf{(ii)} The real field variables
\begin{equation}
\left( \lambda ^{\alpha },\text{ }\lambda _{\dot{\alpha}}|\text{ }\zeta
_{+}^{\alpha },\text{ }\zeta _{+\dot{\alpha}}\right) ,
\end{equation}%
and their conjugate momenta,
\begin{equation}
\left( \mathrm{\nu }_{\alpha }\text{, }\mathrm{\nu }^{\dot{\alpha}}\text{ }|%
\text{ }\mathrm{\xi }_{-\alpha },\text{ }\mathrm{\xi }_{-}^{\dot{\alpha}%
}\right) .
\end{equation}%
are related as,%
\begin{eqnarray}
\mathrm{\nu }_{\alpha } &=&\frac{\delta \mathcal{S}_{FT}}{\delta \left(
\partial _{\tau }\lambda ^{\alpha }\right) },\qquad \mathrm{\nu }^{\dot{%
\alpha}}=\frac{\delta \mathcal{S}_{FT}}{\delta \left( \partial _{\tau
}\lambda _{\dot{\alpha}}\right) },  \notag \\
\mathrm{\xi }_{-\alpha } &=&\frac{\delta \mathcal{S}_{FT}}{\delta \left(
\partial _{\tau }\zeta _{+}^{\alpha }\right) },\qquad \text{ }\mathrm{\xi }%
_{-}^{\dot{\alpha}}=\frac{\delta \mathcal{S}_{FT}}{\delta \left( \partial
_{\tau }\zeta _{+\dot{\alpha}}\right) }.  \label{c1}
\end{eqnarray}%
and parameterize the full phase space of the fermionic twistor like model,%
\begin{equation}
\mathcal{E}_{phase}=\left\{ \lambda ^{\alpha },\text{ }\lambda _{+\dot{\alpha%
}}\text{, }\mathrm{\nu }_{\alpha }\text{, }\mathrm{\nu }^{\dot{\alpha}%
}|\zeta _{+}^{\alpha },\text{ }\zeta _{+}^{\dot{\alpha}},\text{ }\mathrm{\xi
}_{-\alpha }\text{, }\mathrm{\xi }_{-\dot{\alpha}}\right\} \text{,}
\end{equation}%
These variables satisfy the following graded canonical relations,%
\begin{eqnarray}
\left\{ \lambda ^{\alpha },\mathrm{\nu }_{\beta }\right\} _{GPB} &=&\delta
_{\beta }^{\alpha },\qquad \left\{ \lambda _{+\dot{\beta}},\mathrm{\nu }^{%
\dot{\alpha}}\right\} _{GPB}=\delta _{\dot{\beta}}^{\dot{\alpha}}  \notag \\
\left\{ \zeta _{+}^{\alpha },\mathrm{\xi }_{-\beta }\right\} _{GPB}
&=&\delta _{\beta }^{\alpha },\qquad \left\{ \zeta _{+}^{\dot{\alpha}},%
\mathrm{\xi }_{-\dot{\beta}}\right\} _{GPB}=\delta _{\dot{\beta}}^{\dot{%
\alpha}},
\end{eqnarray}%
where the graded Poisson bracket $\left\{ f,\mathrm{g}\right\} _{GPB}$, of
generic functions $f=f\left( \lambda ,\mathrm{\nu }|\zeta _{+},\mathrm{\xi }%
_{-}\right) $ and $\mathrm{g}=\mathrm{g}\left( \lambda ,\mathrm{\nu }|\zeta
_{+},\mathrm{\xi }_{-}\right) $ living on twistor space, is defined as%
\begin{eqnarray}
\left\{ f,\mathrm{g}\right\} _{GPB} &=&\left( \frac{\partial f}{\partial
\lambda ^{\alpha }}\frac{\partial \mathrm{g}}{\partial \mathrm{\nu }_{\alpha
}}-\frac{\partial f}{\partial \mathrm{\nu }_{\alpha }}\frac{\partial \mathrm{%
g}}{\partial \lambda ^{\alpha }}\right) +\left( \frac{\partial f}{\partial
\zeta _{+}^{\alpha }}\frac{\partial \mathrm{g}}{\partial \mathrm{\xi }%
_{-\alpha }}+\frac{\partial f}{\partial \mathrm{\xi }_{-\alpha }}\frac{%
\partial \mathrm{g}}{\partial \zeta _{+}^{\alpha }}\right)  \notag \\
&&+\left( \frac{\partial f}{\partial \lambda _{+\dot{\alpha}}}\frac{\partial
\mathrm{g}}{\partial \mathrm{\nu }^{\dot{\alpha}}}-\frac{\partial f}{%
\partial \mathrm{\nu }^{\dot{\alpha}}}\frac{\partial \mathrm{g}}{\partial
\lambda _{+\dot{\alpha}}}\right) +\left( \frac{\partial f}{\partial \zeta
_{+}^{\dot{\alpha}}}\frac{\partial \mathrm{g}}{\partial \mathrm{\xi }_{-\dot{%
\alpha}}}+\frac{\partial f}{\partial \mathrm{\xi }_{-\dot{\alpha}}}\frac{%
\partial \mathrm{g}}{\partial \zeta _{+}^{\dot{\alpha}}}\right) .
\end{eqnarray}%
For convenience and later use, it is interesting to combine these variables
into super-coordinates as,%
\begin{eqnarray}
z^{a} &=&\left( \lambda ^{\alpha }\text{ }|\text{ }\mathrm{\xi }_{-}^{\dot{%
\alpha}}\right) ,\qquad \qquad \widetilde{z}^{a}=\left( \zeta _{+}^{\alpha }%
\text{ }|\text{ }\mathrm{\nu }^{\dot{\alpha}}\right) ,  \notag \\
\mathrm{w}^{b} &=&\left( \mathrm{\nu }^{\beta }\text{ }|\text{ }\zeta _{+}^{%
\dot{\beta}}\right) ,\qquad \qquad \widetilde{\mathrm{w}}^{b}=\left( \mathrm{%
\xi }_{-}^{\beta }\text{ }|\text{ }\lambda ^{\dot{\beta}}\right) .
\label{exp}
\end{eqnarray}%
With these super variables, the real projective transformations of the
twistor superspace,
\begin{eqnarray}
\lambda ^{\alpha }\quad &\rightarrow &\quad e^{+\varphi }\lambda ^{\alpha
},\qquad \qquad \zeta _{+}^{\alpha }\quad \rightarrow \quad e^{+\theta
}\zeta _{+}^{\alpha },  \notag \\
\mathrm{\nu }_{\alpha }\quad &\rightarrow &\quad e^{-\varphi }\mathrm{\nu }%
_{\alpha },\qquad \qquad \mathrm{\xi }_{-\alpha }\quad \rightarrow \quad
e^{-\theta }\mathrm{\xi }_{-\alpha },  \notag \\
\zeta _{+}^{\dot{\alpha}}\quad &\rightarrow &\quad e^{-\varphi }\zeta _{+}^{%
\dot{\alpha}},\qquad \qquad \lambda _{\dot{\alpha}}\quad \rightarrow \quad
e^{-\theta }\lambda _{\dot{\alpha}},  \label{pra} \\
\mathrm{\xi }_{-\dot{\alpha}}\quad &\rightarrow &\quad e^{+\varphi }\mathrm{%
\xi }_{-\dot{\alpha}},\qquad \qquad \mathrm{\nu }^{\dot{\alpha}}\quad
\rightarrow \quad e^{+\theta }\mathrm{\nu }^{\dot{\alpha}},  \notag
\end{eqnarray}%
combine as as follows,%
\begin{eqnarray}
z^{a}\quad &\rightarrow &\quad e^{+\varphi }z^{a},\qquad \qquad \widetilde{z}%
^{a}\quad \rightarrow \quad e^{+\theta }\widetilde{z}^{a},  \notag \\
\mathrm{w}^{a}\quad &\rightarrow &\quad e^{-\varphi }\mathrm{w}^{a},\qquad
\qquad \widetilde{\mathrm{w}}^{a}\quad \rightarrow \quad e^{-\theta }%
\widetilde{\mathrm{w}}^{a}.  \label{ttr}
\end{eqnarray}%
Note that the parameters $\varphi $ and $\theta $ are a\ priori independent,
unless if the momentum vector,
\begin{equation}
\mathrm{p}_{\dot{\alpha}}^{\alpha }=\lambda ^{\alpha }\widetilde{\lambda }_{%
\dot{\alpha}}
\end{equation}%
is required to be invariant under (\ref{pra}). In this case, we should have
the identity,%
\begin{equation}
\varphi =\theta .  \label{ft}
\end{equation}%
Note also that within the $\left( \Upsilon _{+\alpha }^{\dot{\alpha}},%
{\Large \pi }_{-\dot{\alpha}}^{\alpha }\right) $ sector eq(\ref{sec1}), the
twistor like variables $\left( \lambda ^{\alpha },\text{ }\mathrm{\xi }_{-%
\dot{\alpha}},\text{ }\mathrm{\nu }_{\alpha },\text{ }\zeta _{+}^{\dot{\alpha%
}}\right) $ splits into blocs as follows,
\begin{eqnarray}
z^{a} &=&\left( \lambda ^{\alpha }\text{ }|\text{ }\mathrm{\xi }_{-}^{\dot{%
\alpha}}\right) \text{ \ }\in \text{ \ }RP_{\left( +1,+1,+1,+1\right)
}^{\left( 1|2\right) },  \notag \\
\mathrm{w}^{b} &=&\left( \mathrm{\nu }^{\beta }\text{ }|\text{ }\zeta _{+}^{%
\dot{\beta}}\right) \text{ \ }\in \text{ \ }RP_{\left( -1,-1,-1,-1\right)
}^{\left( 1|2\right) }.  \label{blc}
\end{eqnarray}%
Moreover using the identities (\ref{c1}), the component twistor field action
reads, up to gauge coupling terms type $\mathrm{\xi }_{\mp \dot{\alpha}%
}V\zeta _{\pm }^{\dot{\alpha}}$ to be given later, as follows,%
\begin{eqnarray}
\mathcal{S}_{FT} &=&\frac{c_{1}}{2}\int d\tau \left[ \mathrm{\nu }_{\alpha
}\partial _{\tau }\lambda ^{\alpha }-\lambda ^{\alpha }\left( \partial
_{\tau }\mathrm{\nu }_{\alpha }\right) \right]  \notag \\
&&-\frac{c_{1}^{\prime }}{2}\int d\tau \left[ \mathrm{\nu }^{\dot{\alpha}%
}\partial _{\tau }\lambda _{\dot{\alpha}}-\lambda _{\dot{\alpha}}\left(
\partial _{\tau }\mathrm{\nu }^{\dot{\alpha}}\right) \right]  \notag \\
&&-\frac{c_{2}}{2}\int d\tau \left[ \mathrm{\xi }_{-}^{\dot{\alpha}}\partial
_{\tau }\zeta _{+\dot{\alpha}}+\zeta _{+\dot{\alpha}}\left( \partial _{\tau }%
\mathrm{\xi }_{-}^{\dot{\alpha}}\right) \right]  \label{sft} \\
&&+\frac{c_{2}^{\prime }}{2}\int d\tau \left[ \mathrm{\xi }_{-\alpha
}\partial _{\tau }\zeta _{+}^{\alpha }+\zeta _{+}^{\alpha }\left( \partial
_{\tau }\mathrm{\xi }_{-\alpha }\right) \right] .  \notag
\end{eqnarray}%
The number $c_{1}$, $c_{1}^{\prime }$, $c_{2}$ and $c_{2}^{\prime }$ are
determined by requiring this action to coincide with the action (\ref{x8}),%
\begin{equation}
\mathcal{S}_{FT}\left[ \Upsilon ,{\Large \pi },\widetilde{\Upsilon },%
\widetilde{{\Large \pi }}\right] =\int_{\mathbb{R}}d\tau \mathrm{Tr}\left(
{\Large \pi }_{-\dot{\alpha}}^{\alpha }\frac{\partial }{\partial \tau }%
\Upsilon _{+\alpha }^{\dot{\alpha}}+\widetilde{{\Large \pi }}_{-\dot{\alpha}%
}^{\alpha }\frac{\partial }{\partial \tau }\widetilde{\Upsilon }_{+\alpha }^{%
\dot{\alpha}}\right) .
\end{equation}%
This requirement allows also to determine the fermionic analog of the
Penrose incidence relation (\ref{inc}). Setting,%
\begin{eqnarray}
\mathrm{\nu }_{\alpha } &=&{\Large \pi }_{-\alpha }^{\dot{\alpha}}\zeta _{+%
\dot{\alpha}},  \notag \\
\mathrm{\xi }_{-\dot{\alpha}} &=&{\Large \pi }_{-\dot{\alpha}}^{\alpha
}\lambda _{\alpha },,  \label{finc}
\end{eqnarray}%
and similarly
\begin{eqnarray}
\mathrm{\xi }_{-}^{\alpha } &=&\widetilde{{\Large \pi }}_{-\dot{\alpha}%
}^{\alpha }\lambda ^{\dot{\alpha}},  \notag \\
\mathrm{\nu }^{\dot{\alpha}} &=&\widetilde{{\Large \pi }}_{-\alpha }^{\dot{%
\alpha}}\zeta _{+}^{\alpha }.
\end{eqnarray}%
Then putting back into eq(\ref{sft}), we get for the integrand $\left(
\mathrm{\nu }_{\alpha }\partial _{\tau }\lambda ^{\alpha }+c_{1}\mathrm{\xi }%
_{-\dot{\alpha}}\partial _{\tau }\zeta _{+}^{\dot{\alpha}}\right) $ the
following,%
\begin{equation}
{\Large \pi }_{-\dot{\alpha}}^{\alpha }\zeta _{+}^{\dot{\alpha}}\partial
_{\tau }\lambda _{\alpha }+{\Large \pi }_{-\dot{\alpha}}^{\alpha }\lambda
_{\alpha }\partial _{\tau }\zeta _{+}^{\dot{\alpha}}
\end{equation}%
provided
\begin{equation}
c_{1}=c_{2}=1.
\end{equation}%
A similar result is valid foe the twild sector. Moreover using the
supervariables $z^{a}$ and $\mathrm{w}^{b}$, one can rewrite the previous
twistor field action as follows,%
\begin{eqnarray}
\mathcal{S}_{FT}\left[ z,\mathrm{w,}\widetilde{z},\widetilde{\mathrm{w}}%
\right] &=&\frac{1}{2}\int d\tau \left[ \mathrm{w}^{b}\partial _{\tau
}z^{a}-\left( \partial _{\tau }\mathrm{w}^{b}\right) z^{a}\right] \omega
_{ab}  \notag \\
&&+\frac{1}{2}\int d\tau \left[ \widetilde{\mathrm{w}}^{b}\partial _{\tau }%
\widetilde{z}^{a}-\left( \partial _{\tau }\widetilde{\mathrm{w}}^{b}\right)
\widetilde{z}^{a}\right] \omega _{ab}  \label{stt}
\end{eqnarray}%
with metric $\omega _{ab}$ given by%
\begin{equation}
\omega _{ab}=\left(
\begin{array}{cc}
\epsilon _{\alpha \beta } & 0 \\
0 & \epsilon _{\dot{\alpha}\dot{\beta}}%
\end{array}%
\right) ,\qquad \omega ^{ba}=\left(
\begin{array}{cc}
\epsilon ^{\beta \alpha } & 0 \\
0 & \epsilon ^{\dot{\beta}\dot{\alpha}}%
\end{array}%
\right) .
\end{equation}%
From these eqs, one reads a set of features; in particular the following:
\newline
(\textbf{1}) the variable $\mathrm{w}_{a}=\omega _{ab}\mathrm{w}^{b}$ is the
conjugate momentum of $z^{a}$ and so we have amongst others,%
\begin{eqnarray}
\left\{ z^{a},\mathrm{w}^{b}\right\} _{GPB} &=&\omega ^{ba},\qquad  \notag \\
\left\{ z^{a},z^{b}\right\} _{GPB} &=&\left\{ \mathrm{w}_{a},\mathrm{w}%
_{b}\right\} _{GPB}=0,
\end{eqnarray}%
where $\omega ^{ab}$ stands for the inverse of $\omega _{ba}$; that is $%
\omega ^{ab}\omega _{bc}=\delta _{c}^{a}$. We also have for generic
functions $f=f\left( z,\mathrm{w,}\widetilde{z},\widetilde{\mathrm{w}}%
\right) $ and $\mathrm{g}=\mathrm{g}\left( z,\mathrm{w,}\widetilde{z},%
\widetilde{\mathrm{w}}\right) $, the following
\begin{eqnarray}
\left\{ f,\mathrm{g}\right\} _{GPB} &=&\omega ^{ba}\left( \frac{\partial f}{%
\partial z^{a}}\frac{\partial \mathrm{g}}{\partial \mathrm{w}^{b}}-\left(
-\right) ^{\left\vert a\right\vert .\left\vert b\right\vert }\frac{\partial f%
}{\partial \mathrm{w}^{b}}\frac{\partial \mathrm{g}}{\partial z^{a}}\right)
\notag \\
&&+\omega ^{ba}\left( \frac{\partial f}{\partial \widetilde{z}^{a}}\frac{%
\partial \mathrm{g}}{\partial \widetilde{\mathrm{w}}^{b}}-\left( -\right)
^{\left\vert a\right\vert .\left\vert b\right\vert }\frac{\partial f}{%
\partial \widetilde{\mathrm{w}}^{b}}\frac{\partial \mathrm{g}}{\partial
\widetilde{z}^{a}}\right) ,
\end{eqnarray}%
where $\left\vert a\right\vert \equiv grad\left( a\right) $ with,%
\begin{equation}
grad\left( a\right) =\left\{
\begin{array}{c}
grad\left( \alpha \right) =0 \\
grad\left( \pm \dot{\alpha}\right) =1%
\end{array}%
\right.
\end{equation}%
(\textbf{2}) Eq(\ref{stt}) is manifestly invariant under global projective
transformations (\ref{ttr}) and $SL\left( 2|2,R\right) $ supergroup of the
twistor supermanifold $RP^{\left( 1|2\right) }$.

\subsection{Quantum constraints}

\qquad Promoting the projective transformations (\ref{pra}-\ref{ttr}) to
local ones $\varphi =\varphi \left( \tau \right) $; $\theta =\theta \left(
\tau \right) $ that is $\partial _{\tau }\varphi \neq 0,$ $\partial _{\tau
}\theta \neq 0$, then gauge invariance of the field action (\ref{stt})
requires implementation of real gauge fields,
\begin{equation}
\mathrm{V}_{\tau }\left( \tau \right) \equiv \mathrm{V}\left( \tau \right)
,\qquad \widetilde{\mathrm{V}}_{\tau }\left( \tau \right) \equiv \widetilde{%
\mathrm{V}}\left( \tau \right)
\end{equation}%
with gauge transformations
\begin{eqnarray*}
\mathrm{V}^{\prime } &=&\mathrm{V}+\partial _{\tau }\varphi , \\
\widetilde{\mathrm{V}}^{\prime } &=&\widetilde{\mathrm{V}}+\partial _{\tau
}\theta .
\end{eqnarray*}
Thus, the action (\ref{stt}) extends as,%
\begin{eqnarray}
\mathcal{S}_{FT}\left[ z,\mathrm{w,V,}\widetilde{z},\widetilde{\mathrm{w}},%
\widetilde{\mathrm{V}}\right] &=&\frac{1}{2}\left( \int d\tau \left[ \mathrm{%
w}^{b}D_{\tau }z^{a}-\left( D_{\tau }\mathrm{w}^{b}\right) z^{a}\right]
\omega _{ab}\right)  \notag \\
&&+\frac{1}{2}\left( \int d\tau \left[ \widetilde{\mathrm{w}}^{b}\widetilde{D%
}_{\tau }\widetilde{z}^{a}-\left( \widetilde{D}_{\tau }\widetilde{\mathrm{w}}%
^{b}\right) \widetilde{z}^{a}\right] \omega _{ab}\right) \\
&&+\int d\tau 2h\mathrm{V}+\int d\tau 2\widetilde{h}\widetilde{\mathrm{V}}
\notag
\end{eqnarray}%
where the gauge covariant derivatives are given by,
\begin{equation}
D_{\tau }z^{a}=\left( \partial _{\tau }-\mathrm{V}\right) z^{a},\qquad
D_{\tau }\mathrm{w}^{b}=\left( \partial _{\tau }+\mathrm{V}\right) \mathrm{w}%
^{b}
\end{equation}
and
\begin{equation}
\widetilde{D}_{\tau }\widetilde{z}^{a}=\left( \partial _{\tau }-\widetilde{%
\mathrm{V}}\right) \widetilde{z}^{a},\qquad \widetilde{D}_{\tau }\widetilde{%
\mathrm{w}}^{b}=\left( \partial _{\tau }+\widetilde{\mathrm{V}}\right)
\widetilde{\mathrm{w}}^{b}
\end{equation}%
The constants $h$ and $\widetilde{h}$\ are a priori real coupling constants;
they take discrete values at quantum level. Since $\mathrm{V}$\ and $%
\widetilde{\mathrm{V}}$\ are non propagating fields, their elimination by
using the equations of motion
\begin{equation*}
\frac{\delta \mathcal{S}_{FT}}{\delta \mathrm{V}}=0,\qquad \frac{\delta
\mathcal{S}_{FT}}{\delta \widetilde{\mathrm{V}}}=0
\end{equation*}
and lead to the following constraint eqs,%
\begin{eqnarray}
\omega _{ab}\mathrm{w}^{b}z^{a}-2h &=&0  \notag \\
\omega _{ab}\widetilde{\mathrm{w}}^{b}\widetilde{z}^{a}-2\widetilde{h} &=&0
\label{tcs}
\end{eqnarray}%
or equivalently by using (\ref{exp}),%
\begin{eqnarray}
\left( \mathrm{\nu }_{\alpha }\lambda ^{\alpha }-\zeta _{+\dot{\alpha}}%
\mathrm{\xi }_{-}^{\dot{\alpha}}\right) -2h &=&0  \notag \\
\left( \mathrm{\nu }_{\dot{\alpha}}\lambda ^{\dot{\alpha}}-\zeta _{+\alpha }%
\mathrm{\xi }_{-}^{\alpha }\right) -2\widetilde{h} &=&0
\end{eqnarray}%
Note that by substituting $\mathrm{\nu }_{\alpha }$ and $\mathrm{\xi }_{-}^{%
\dot{\alpha}}$ and similarly $\mathrm{\nu }_{\dot{\alpha}}$ and $\mathrm{\xi
}_{-}^{\alpha }$ by the relations (\ref{finc}) associated with helicity zero
particles ($h=0$, $\widetilde{h}=0$), one finds that $\left( \mathrm{\nu }%
_{\alpha }\lambda ^{\alpha }-\zeta _{+\dot{\alpha}}\mathrm{\xi }_{-}^{\dot{%
\alpha}}\right) $ and $\left( \mathrm{\nu }_{\dot{\alpha}}\lambda ^{\dot{%
\alpha}}-\zeta _{+\alpha }\mathrm{\xi }_{-}^{\alpha }\right) $ vanish
identically. Note also that eq(\ref{tcs}) can be also rewritten as,%
\begin{eqnarray}
\omega _{ab}\left( \mathrm{w}^{b}z^{a}+\left( -\right) ^{\left\vert
a\right\vert .\left\vert b\right\vert }z^{a}\mathrm{w}^{b}\right) -4h &=&0,
\notag \\
\omega _{ab}\left( \widetilde{\mathrm{w}}^{b}\widetilde{z}^{a}+\left(
-\right) ^{\left\vert a\right\vert .\left\vert b\right\vert }\widetilde{z}%
^{a}\widetilde{\mathrm{w}}^{b}\right) -4\widetilde{h} &=&0  \label{gcd}
\end{eqnarray}%
Quantum mechanically, the graded bracket $\left\{ z^{a},\mathrm{w}%
^{b}\right\} _{GPB}=\omega ^{ba}$ of classical twistor model gets replaced
by a graded commutator,%
\begin{equation}
\left[ Z^{a},\mathrm{W}^{b}\right\} =Z^{a}\mathrm{W}^{b}+\left( -\right)
^{\left\vert a\right\vert .\left\vert b\right\vert }\mathrm{W}%
^{b}Z^{a}=\omega ^{ba}.  \label{zw}
\end{equation}%
A similar relation is valid for the twild variables. Eq(\ref{zw}) and its
twild analog may be solved like,
\begin{eqnarray}
Z^{a} &=&z^{a},\qquad \mathrm{W}^{b}=\left( -\right) ^{\left\vert
a\right\vert .\left\vert b\right\vert }\omega ^{ba}\frac{\partial }{\partial
z^{a}},  \notag \\
\widetilde{Z}^{a} &=&\widetilde{z}^{a},\qquad \widetilde{\mathrm{W}}%
^{b}=\left( -\right) ^{\left\vert a\right\vert .\left\vert b\right\vert
}\omega ^{ba}\frac{\partial }{\partial \widetilde{z}^{a}}.
\end{eqnarray}%
At quantum level, the classical constraint eqs(\ref{gcd}) are replaced by
operators acting on states $\mid \Psi _{h,\widetilde{h}}>$ of the Hilbert
space of the fermionic twistor like system. More precisely, we have,%
\begin{eqnarray}
\omega _{ab}\left( \mathrm{W}^{b}Z^{a}+\left( -\right) ^{\left\vert
a\right\vert .\left\vert b\right\vert }Z^{a}\mathrm{W}^{b}\right) &\mid
&\Psi _{h,\widetilde{h}}>=4h\mid \Psi _{h,\widetilde{h}}>,  \notag \\
\omega _{ab}\left( \widetilde{\mathrm{W}}^{b}\widetilde{Z}^{a}+\left(
-\right) ^{\left\vert a\right\vert .\left\vert b\right\vert }\widetilde{Z}%
^{a}\widetilde{\mathrm{W}}^{b}\right) &\mid &\Psi _{h,\widetilde{h}}>=4%
\widetilde{h}\mid \Psi _{h,\widetilde{h}}>,
\end{eqnarray}%
or equivalently by using $\Psi \left( z,\widetilde{z}\right) =<z,\widetilde{z%
}\mid \Psi >,$
\begin{eqnarray}
\omega _{ab}\left( -\right) ^{\left\vert a\right\vert .\left\vert
b\right\vert }\left[ \omega ^{ba}+2\left( -\right) ^{\left\vert c\right\vert
.\left\vert b\right\vert }\omega ^{bc}z^{a}\frac{\partial }{\partial z^{c}}%
\right] \Psi _{h,\widetilde{h}}\left( z,\widetilde{z}\right) &=&4h\Psi _{h,%
\widetilde{h}}\left( z,\widetilde{z}\right) ,  \notag \\
\omega _{ab}\left( -\right) ^{\left\vert a\right\vert .\left\vert
b\right\vert }\left[ \omega ^{ba}+2\left( -\right) ^{\left\vert c\right\vert
.\left\vert b\right\vert }\omega ^{bc}\widetilde{z}^{a}\frac{\partial }{%
\partial \widetilde{z}^{c}}\right] \Psi _{h,\widetilde{h}}\left( z,%
\widetilde{z}\right) &=&4h\Psi _{h,\widetilde{h}}\left( z,\widetilde{z}%
\right) .
\end{eqnarray}%
The first term of this relation $\sum \left( -\right) ^{\left\vert
a\right\vert .\left\vert b\right\vert }\omega _{ab}\omega ^{ba}$ is
proportional to the supertrace of $4\times 4$ matrix,%
\begin{equation}
Str\left( \omega _{ab}\omega ^{ba}\right) =Str\left(
\begin{array}{cc}
I & 0 \\
0 & I%
\end{array}%
\right) ,
\end{equation}%
which vanishes identically. So one is left with the following,%
\begin{eqnarray}
\left( \lambda ^{\alpha }\frac{\partial }{\partial \lambda ^{\alpha }}+%
\mathrm{\xi }_{-}^{\dot{\alpha}}\frac{\partial }{\partial \mathrm{\xi }_{-}^{%
\dot{\alpha}}}\right) \Psi _{h,\widetilde{h}}\left( z,\widetilde{z}\right)
&=&2h\Psi _{h,\widetilde{h}}\left( z,\widetilde{z}\right) ,  \notag \\
\left( \mathrm{\nu }^{\dot{\alpha}}\frac{\partial }{\partial \mathrm{\nu }^{%
\dot{\alpha}}}+\zeta _{+}^{\alpha }\frac{\partial }{\partial \zeta
_{+}^{\alpha }}\right) \Psi _{h,\widetilde{h}}\left( z,\widetilde{z}\right)
&=&2\widetilde{h}\Psi _{h,\widetilde{h}}\left( z,\widetilde{z}\right)
\end{eqnarray}%
showing that twistor space wave functions $\Psi _{h,\widetilde{h}}(z,%
\widetilde{z})$ are homogeneous superfunctions of degree $\left( 2h,2%
\widetilde{h}\right) $,%
\begin{equation}
\Psi (e^{+\varphi }z,e^{+\theta }\widetilde{z})=e^{+\left( 2h+2\widetilde{h}%
\right) \varphi }\Psi (z,\widetilde{z}).  \label{trs}
\end{equation}%
Charge quantization of scaling symmetry requires $2h$ and $2\widetilde{h}$\
to be integer,%
\begin{equation}
2h\text{ \ \ }\in \text{ \ \ }\mathbb{Z},\qquad 2\widetilde{h}\text{ \ \ }%
\in \text{ \ \ }\mathbb{Z},
\end{equation}%
in agreement with local properties of quantum field theory requiring
helicity to take half integer values.\ Moreover, wave function $\Psi _{h,%
\widetilde{h}}(z,\widetilde{z})=\Psi _{h,\widetilde{h}}\left( \lambda
^{\alpha },\mathrm{\nu }^{\dot{\alpha}}|\zeta _{+}^{\alpha },\mathrm{\xi }_{-%
\dot{\alpha}}\right) $ can be viewed as superfields with expansions type (%
\ref{ex}). We have%
\begin{eqnarray}
\Psi (z,\widetilde{z}) &=&{\normalsize \phi }+\zeta _{+}^{\alpha }%
{\normalsize \psi }_{-\alpha }+\mathrm{\xi }_{-\dot{\alpha}}{\normalsize %
\psi }_{+}^{\dot{\alpha}}+\zeta _{+}^{2}{\normalsize F}_{--}+\mathrm{\xi }%
_{-}^{2}{\normalsize F}_{++}+\zeta _{+}^{\alpha }\mathrm{\xi }_{-\dot{\alpha}%
}{\normalsize A}_{\alpha }^{\dot{\alpha}}  \notag \\
&&+\mathrm{\xi }_{-}^{2}\zeta _{+}^{\alpha }{\normalsize \chi }_{+\alpha
}+\zeta _{+}^{2}\mathrm{\xi }_{-\dot{\alpha}}{\normalsize \chi }_{-}^{\dot{%
\alpha}}+\zeta _{+}^{2}\mathrm{\xi }_{-}^{2}{\normalsize \Delta }
\label{sup}
\end{eqnarray}%
where the component fields are as $\phi =\phi _{h,\widetilde{h}}\left(
\lambda ^{\alpha },\mathrm{\nu }^{\dot{\alpha}}\right) $ and so on. Using
the transformation laws (\ref{ttr},\ref{trs}), one can write down the
corresponding ones for the component fields.

\subsection{Comments}

\qquad Here we want to make three comments regarding the study developed
above:\newline
(\textbf{1}) The analysis done for the $\mathbb{R}^{\left( 2,2\right) }$
case with group $SO\left( 2,2\right) $\ can be naturally extended for
Minkowski $\mathbb{R}^{\left( 1,3\right) }$ and Euclidean $\mathbb{R}^{4}$
geometries with rotation groups $SO\left( 1,3\right) $ and $SO\left(
4\right) $ respectively. For the case of $\mathbb{R}^{\left( 1,3\right) }$,
spinors are in $SL\left( 2,C\right) $ and the correspondence between generic
superfield (\ref{sup}) and 4D $\mathcal{N}=1$ supersymmetry representations
is manifest.\newline
(\textbf{2}) Target space supersymmetry is generated by the following
superspace translations,%
\begin{eqnarray}
\delta \zeta _{+}^{\alpha } &\sim &\varepsilon _{+}^{\alpha },  \notag \\
\delta \mathrm{\xi }_{-\dot{\alpha}} &\sim &\varepsilon _{-\dot{\alpha}},
\notag \\
\delta \lambda ^{\alpha } &\sim &\zeta _{+}^{\alpha }\varepsilon _{-\dot{%
\alpha}}\mathrm{\nu }^{\dot{\alpha}} \\
\delta \mathrm{\nu }^{\dot{\alpha}} &\sim &\mathrm{\xi }_{-}^{\dot{\alpha}%
}\varepsilon _{+}^{\alpha }\lambda _{\alpha }  \notag
\end{eqnarray}%
where $\varepsilon _{+}^{\alpha }$ and $\varepsilon _{-\dot{\alpha}}$\ are
the target space supersymmetric parameters. These transformations are
motivated by group covariance and dimensional arguments. Putting these naive
transformations back into eq(\ref{sup}), we can get the supersymmetric
transformations of the target space field variables. For the leading
component fields, we have the following transformations,%
\begin{eqnarray}
\delta {\normalsize \phi } &\sim &\varepsilon _{+}^{\alpha }{\normalsize %
\psi }_{-\alpha }+\varepsilon _{-\dot{\alpha}}{\normalsize \psi }_{+}^{\dot{%
\alpha}}  \notag \\
\delta {\normalsize \psi }_{-\alpha } &\sim &2\varepsilon _{+\alpha }%
{\normalsize F}_{--}+\varepsilon _{-\dot{\alpha}}\left( \mathrm{\nu }^{\dot{%
\alpha}}\frac{\partial }{\partial \lambda ^{\alpha }}{\normalsize \phi
-A_{\alpha }^{\dot{\alpha}}}\right) {\normalsize ,} \\
\delta {\normalsize \psi }_{+}^{\dot{\alpha}} &\sim &2\varepsilon _{-}^{\dot{%
\alpha}}{\normalsize F}_{++}+\varepsilon _{+}^{\alpha }\left( \lambda
_{\alpha }\frac{\partial }{\partial \mathrm{\nu }_{\dot{\alpha}}}%
{\normalsize \phi +A_{\alpha }^{\dot{\alpha}}}\right) ,  \notag
\end{eqnarray}%
and so on. Now performing a second supersymmetric transformation on $\delta
{\normalsize \phi }$ with supersymmetric parameters $\varepsilon _{+\alpha
}^{\prime }$ and $\varepsilon _{-\dot{\alpha}}^{\prime }$, we have $\delta
^{\prime }\delta {\normalsize \phi \sim }$ $\varepsilon _{+}^{\alpha }\delta
^{\prime }{\normalsize \psi }_{-\alpha }+\varepsilon _{-\dot{\alpha}}\delta
^{\prime }{\normalsize \psi }_{+}^{\dot{\alpha}}$. Then replacing $\delta
^{\prime }{\normalsize \psi }_{-\alpha }$ and $\delta ^{\prime }{\normalsize %
\psi }_{+}^{\dot{\alpha}}$ by their expressions, we get\
\begin{eqnarray}
\delta ^{\prime }\delta {\normalsize \phi } &\sim &\text{ }\left[
2\varepsilon _{+}^{\alpha }\varepsilon _{+\alpha }^{\prime }{\normalsize F}%
_{--}+\varepsilon _{+}^{\alpha }\varepsilon _{-\dot{\alpha}}^{\prime }\left(
\mathrm{\nu }^{\dot{\alpha}}\frac{\partial }{\partial \lambda ^{\alpha }}%
{\normalsize \phi -A_{\alpha }^{\dot{\alpha}}}\right) \right]  \notag \\
&&+\left[ 2\varepsilon _{-\dot{\alpha}}\varepsilon _{-}^{\prime \dot{\alpha}}%
{\normalsize F}_{++}+\varepsilon _{-\dot{\alpha}}\varepsilon _{+}^{\prime
\alpha }\left( \lambda _{\alpha }\frac{\partial }{\partial \mathrm{\nu }_{%
\dot{\alpha}}}{\normalsize \phi +A_{\alpha }^{\dot{\alpha}}}\right) \right]
\end{eqnarray}%
From this relation, one can compute the commutator $\left[ \delta ^{\prime
},\delta \right] {\normalsize \phi }$, we find%
\begin{eqnarray}
\left[ \delta ^{\prime },\delta \right] {\normalsize \phi } &\sim &\text{ }%
\left( \varepsilon _{+}^{\alpha }\varepsilon _{-\dot{\alpha}}^{\prime
}-\varepsilon _{+}^{\alpha \prime }\varepsilon _{-\dot{\alpha}}\right)
\mathrm{\nu }^{\dot{\alpha}}\frac{\partial }{\partial \lambda ^{\alpha }}%
{\normalsize \phi }  \notag \\
&&+\left( \varepsilon _{-\dot{\alpha}}\varepsilon _{+}^{\prime \alpha
}-\varepsilon _{-\dot{\alpha}}^{\prime }\varepsilon _{+}^{\alpha }\right)
\lambda _{\alpha }\frac{\partial }{\partial \mathrm{\nu }_{\dot{\alpha}}}%
{\normalsize \phi }  \notag \\
&&+2\left( \varepsilon _{+}^{\alpha }\varepsilon _{+\alpha }^{\prime
}-\varepsilon _{+}^{\alpha \prime }\varepsilon _{+\alpha }\right)
{\normalsize F}_{--}+2\left( \varepsilon _{-\dot{\alpha}}\varepsilon
_{-}^{\prime \dot{\alpha}}-\varepsilon _{-\dot{\alpha}}^{\prime }\varepsilon
_{-}^{\dot{\alpha}}\right) {\normalsize F}_{++} \\
&&\left[ \left( \varepsilon _{-\dot{\alpha}}\varepsilon _{+}^{\prime \alpha }%
{\normalsize +\varepsilon _{+}^{\alpha \prime }\varepsilon _{-\dot{\alpha}}}%
\right) -\left( \varepsilon _{-\dot{\alpha}}^{\prime }\varepsilon
_{+}^{\alpha }{\normalsize +\varepsilon _{+}^{\alpha }\varepsilon _{-\dot{%
\alpha}}^{\prime }}\right) \right] {\normalsize A_{\alpha }^{\dot{\alpha}}.}
\notag
\end{eqnarray}%
But this relation may be simplified by using the following identities,%
\begin{eqnarray}
\varepsilon _{+}^{\alpha }\varepsilon _{+\alpha }^{\prime } &=&\varepsilon
_{+}^{\alpha \prime }\varepsilon _{+\alpha },  \notag \\
\varepsilon _{-\dot{\alpha}}\varepsilon _{-}^{\prime \dot{\alpha}}
&=&\varepsilon _{-\dot{\alpha}}^{\prime }\varepsilon _{-}^{\dot{\alpha}},
\notag \\
\varepsilon _{-\dot{\alpha}}\varepsilon _{+}^{\prime \alpha } &{\normalsize =%
}&{\normalsize -\varepsilon _{+}^{\alpha \prime }\varepsilon _{-\dot{\alpha}%
},} \\
\varepsilon _{-\dot{\alpha}}^{\prime }\varepsilon _{+}^{\alpha } &=&-%
{\normalsize \varepsilon _{+}^{\alpha }\varepsilon _{-\dot{\alpha}}^{\prime }%
}  \notag \\
\left( \varepsilon _{+}^{\alpha }\varepsilon _{-\dot{\alpha}}^{\prime
}-\varepsilon _{+}^{\alpha \prime }\varepsilon _{-\dot{\alpha}}\right)
&=&-\left( \varepsilon _{-\dot{\alpha}}^{\prime }\varepsilon _{+}^{\alpha
}-\varepsilon _{-\dot{\alpha}}\varepsilon _{+}^{\prime \alpha }\right)
{\normalsize .}  \notag
\end{eqnarray}%
Taking into account these features, one ends with the result,%
\begin{equation}
\left[ \delta _{\varepsilon ^{\prime }},\delta _{\varepsilon }\right] \text{
}{\normalsize \phi \sim }\text{ }-\left( \varepsilon _{-\dot{\alpha}%
}^{\prime }\varepsilon _{+}^{\alpha }-\varepsilon _{-\dot{\alpha}%
}\varepsilon _{+}^{\prime \alpha }\right) {\normalsize P_{\alpha }^{\dot{%
\alpha}}}\text{ }{\normalsize \phi ,}
\end{equation}%
with%
\begin{equation}
{\normalsize P_{\alpha }^{\dot{\alpha}}}=\mathrm{\nu }^{\dot{\alpha}}\frac{%
\partial }{\partial \lambda ^{\alpha }}+\lambda _{\alpha }\frac{\partial }{%
\partial \mathrm{\nu }_{\dot{\alpha}}}.
\end{equation}%
This supersymmetric transformation should be hold also by the remaining
component fields. As such target space supersymmetry can be defined in the
twistor like space described above as follows,%
\begin{eqnarray}
\left\{ \mathrm{D}_{-\alpha },\mathrm{D}_{+}^{\dot{\alpha}}\right\} \text{ }
&=&2\text{ \textrm{P}}_{\alpha }^{\dot{\alpha}},  \notag \\
\left[ \text{\textrm{P}}_{\alpha }^{\dot{\alpha}},\mathrm{D}_{-\alpha }%
\right] \text{ } &=&\text{ }\left[ \text{\textrm{P}}_{\alpha }^{\dot{\alpha}%
},\mathrm{D}_{+}^{\dot{\alpha}}\right] \text{ }=0.  \label{50}
\end{eqnarray}%
Using the twistor like variables $\left( \lambda _{\alpha },\mathrm{\nu }_{%
\dot{\alpha}},\zeta _{+}^{\alpha },\mathrm{\xi }_{-\dot{\alpha}}\right) $,
this superalgera can be realized as,
\begin{eqnarray}
\mathrm{D}_{+}^{\dot{\alpha}} &=&\frac{\partial }{\partial \mathrm{\xi }_{-%
\dot{\alpha}}}+\zeta _{+}^{\alpha }\lambda _{\alpha }\frac{\partial }{%
\partial \mathrm{\nu }_{\dot{\alpha}}},  \notag \\
\mathrm{D}_{-\alpha } &=&\frac{\partial }{\partial \zeta _{+}^{\alpha }}+%
\mathrm{\xi }_{-\dot{\alpha}}\mathrm{\nu }^{\dot{\alpha}}\frac{\partial }{%
\partial \lambda ^{\alpha }},  \label{51} \\
\text{\textrm{P}}_{\alpha }^{\dot{\alpha}} &=&\mathrm{\nu }^{\dot{\alpha}}%
\frac{\partial }{\partial \lambda ^{\alpha }}+\lambda _{\alpha }\frac{%
\partial }{\partial \mathrm{\nu }_{\dot{\alpha}}}.  \notag
\end{eqnarray}%
Note that the above realization is not the unique one in twistor like space;
we have in addition two more "chiral" ones. The first one corresponds to,%
\begin{eqnarray}
\mathrm{D}_{+}^{\dot{\alpha}} &=&\frac{\partial }{\partial \mathrm{\xi }_{-%
\dot{\alpha}}},  \notag \\
\mathrm{D}_{-\alpha } &=&\frac{\partial }{\partial \zeta _{+}^{\alpha }}+2%
\mathrm{\xi }_{-\dot{\alpha}}\mathrm{\nu }^{\dot{\alpha}}\frac{\partial }{%
\partial \lambda ^{\alpha }},  \label{h} \\
\text{\textrm{P}}_{\alpha }^{\dot{\alpha}} &=&\mathrm{\nu }^{\dot{\alpha}}%
\frac{\partial }{\partial \lambda ^{\alpha }}.  \notag
\end{eqnarray}%
and the second one is given by the representation,%
\begin{eqnarray}
\mathrm{D}_{+}^{\dot{\alpha}} &=&\frac{\partial }{\partial \mathrm{\xi }_{-%
\dot{\alpha}}}+2\zeta _{+}^{\alpha }\lambda _{\alpha }\frac{\partial }{%
\partial \mathrm{\nu }_{\dot{\alpha}}},  \notag \\
\mathrm{D}_{-\alpha } &=&\frac{\partial }{\partial \zeta _{+}^{\alpha }},
\label{ha} \\
\text{\textrm{P}}_{\alpha }^{\dot{\alpha}} &=&\lambda _{\alpha }\frac{%
\partial }{\partial \mathrm{\nu }_{\dot{\alpha}}}.  \notag
\end{eqnarray}%
Note also that in chiral eqs, the original energy momentum vector \textrm{P}$%
_{\alpha }^{\dot{\alpha}}=\left( \mathrm{\nu }^{\dot{\alpha}}\frac{\partial
}{\partial \lambda ^{\alpha }}+\lambda _{\alpha }\frac{\partial }{\partial
\mathrm{\nu }_{\dot{\alpha}}}\right) $ of eqs(\ref{51}) has been split into
two parts. Each part appears in one of the two chiral realizations; i.e $%
\mathrm{\nu }^{\dot{\alpha}}\frac{\partial }{\partial \lambda ^{\alpha }}$
in (\ref{h}) and $\lambda _{\alpha }\frac{\partial }{\partial \mathrm{\nu }_{%
\dot{\alpha}}}$ in (\ref{ha}).\newline
(\textbf{3}) The twistor like projective transformations (\ref{pra}) can be
restricted by using the identity (\ref{ft}). In this case, the gauge fields $%
\mathrm{V}\left( \tau \right) $ and $\widetilde{\mathrm{V}}_{\tau }\left(
\tau \right) $ get identified and so the field action reduces to,%
\begin{eqnarray}
\mathcal{S}_{FT}\left[ z,\mathrm{w,}\widetilde{z},\widetilde{\mathrm{w}},%
\mathrm{V}\right] &=&\frac{1}{2}\left( \int d\tau \left[ \mathrm{w}%
^{b}D_{\tau }z^{a}-\left( D_{\tau }\mathrm{w}^{b}\right) z^{a}\right] \omega
_{ab}\right)  \notag \\
&&+\frac{1}{2}\left( \int d\tau \left[ \widetilde{\mathrm{w}}^{b}D_{\tau }%
\widetilde{z}^{a}-\left( D_{\tau }\widetilde{\mathrm{w}}^{b}\right)
\widetilde{z}^{a}\right] \omega _{ab}\right) \\
&&+\int d\tau 2h\mathrm{V}.  \notag
\end{eqnarray}%
By eliminating $\mathrm{V}$\ through its equation of motion, we get the
following constraint eq,%
\begin{equation}
\omega _{ab}\mathrm{w}^{b}z^{a}+\omega _{ab}\widetilde{\mathrm{w}}^{b}%
\widetilde{z}^{a}-2h=0
\end{equation}%
or equivalently by using (\ref{exp}),%
\begin{equation}
\left( \mathrm{\nu }_{\alpha }\lambda ^{\alpha }-\zeta _{+\dot{\alpha}}%
\mathrm{\xi }_{-}^{\dot{\alpha}}\right) -\left( \mathrm{\nu }_{\dot{\alpha}%
}\lambda ^{\dot{\alpha}}-\zeta _{+\alpha }\mathrm{\xi }_{-}^{\alpha }\right)
-2h=0
\end{equation}%
This constraint eq can be also rewritten as,
\begin{equation}
\omega _{ab}\left[ \left( \mathrm{w}^{b}z^{a}+\left( -\right) ^{\left\vert
a\right\vert .\left\vert b\right\vert }z^{a}\mathrm{w}^{b}\right) +\left(
\widetilde{\mathrm{w}}^{b}\widetilde{z}^{a}+\left( -\right) ^{\left\vert
a\right\vert .\left\vert b\right\vert }\widetilde{z}^{a}\widetilde{\mathrm{w}%
}^{b}\right) \right] -4h=0  \label{gcc}
\end{equation}%
and at quantum level it gets promoted to,%
\begin{equation}
\omega _{ab}\left[ \left( \mathrm{W}^{b}Z^{a}+\left( -\right) ^{\left\vert
a\right\vert .\left\vert b\right\vert }Z^{a}\mathrm{W}^{b}\right) +%
\widetilde{\mathrm{W}}^{b}\widetilde{Z}^{a}+\left( -\right) ^{\left\vert
a\right\vert .\left\vert b\right\vert }\widetilde{Z}^{a}\widetilde{\mathrm{W}%
}^{b}\right] \left\vert \Psi \right\rangle =4h\left\vert \Psi \right\rangle .
\label{gqc}
\end{equation}%
In differential representation $Z=z$, $\mathrm{W}\sim \frac{\partial }{%
\partial z}$ and so on, this constraint reads as,%
\begin{equation}
\omega _{ab}\left( -\right) ^{\left\vert a\right\vert .\left\vert
b\right\vert }\left[ \left( -\right) ^{\left\vert c\right\vert .\left\vert
b\right\vert }\omega ^{bc}z^{a}\frac{\partial }{\partial z^{c}}+\left(
-\right) ^{\left\vert c\right\vert .\left\vert b\right\vert }\omega ^{bc}%
\widetilde{z}^{a}\frac{\partial }{\partial \widetilde{z}^{c}}\right] \Psi
\left( z,\widetilde{z}\right) =2h\Psi \left( z,\widetilde{z}\right) .
\end{equation}%
where we have used the identity $\sum \left( -\right) ^{\left\vert
a\right\vert .\left\vert b\right\vert }\omega _{ab}\omega ^{ba}=0$. So one
is left with the following,%
\begin{equation}
\left( \lambda ^{\alpha }\frac{\partial }{\partial \lambda ^{\alpha }}+%
\mathrm{\nu }^{\dot{\alpha}}\frac{\partial }{\partial \mathrm{\nu }^{\dot{%
\alpha}}}+\zeta _{+}^{\alpha }\frac{\partial }{\partial \zeta _{+}^{\alpha }}%
+\mathrm{\xi }_{-}^{\dot{\alpha}}\frac{\partial }{\partial \mathrm{\xi }%
_{-}^{\dot{\alpha}}}\right) \Psi \left( z,\widetilde{z}\right) =2h\Psi
\left( z,\widetilde{z}\right)
\end{equation}%
showing that twistor space wave functions $\Psi (z,\widetilde{z})$ are
homogeneous superfunctions of degree $2h$, $\Psi (e^{+\varphi }z,e^{+\varphi
}\widetilde{z})=e^{+2h\varphi }\Psi (z)$. Charge quantization requires $2h$
to be integer.

\section{Conclusion and discussion}

\qquad In this paper, we have developed a pure fermionic twistor like model
capturing in a remarkable way the link between world line (world sheet)
supersymmetry and target space one. This model describes a field theoretic
system based on solving the nilpotency property
\begin{equation}
\mathrm{\Upsilon }_{\pm }^{2}=\epsilon _{\dot{\alpha}\dot{\beta}}\epsilon
^{\alpha \beta }\mathrm{\Upsilon }_{\pm \alpha }^{\dot{\alpha}}\mathrm{%
\Upsilon }_{\pm \beta }^{\dot{\beta}}=0,\qquad \alpha ,\dot{\alpha}=1,2,
\label{con1}
\end{equation}%
of world line fermions type $\mathrm{\Upsilon }_{\pm \alpha }^{\dot{\alpha}}=%
\mathrm{\Upsilon }_{\pm \alpha }^{\dot{\alpha}}\left( \tau \right) $ and can
be viewed as the analogue of the standard Penrose model for a massless
particle moving in $4D$ space time $\mathbb{R}^{\left( d,4-d\right) }$ with $%
d=2,3,4$. \newline
The construction involves three main levels: \newline
(\textbf{1}) Start with the world line field action $\mathcal{S}_{F}=%
\mathcal{S}_{F}\left[ \mathrm{\Upsilon }\right] $ of a fermionic particle
parameterized by the world line fields $\mathrm{\Upsilon }_{\pm \alpha }^{%
\dot{\alpha}}$ and $\widetilde{\mathrm{\Upsilon }}_{\pm \alpha }^{\dot{\alpha%
}}$. The latter fields ($\widetilde{\mathrm{\Upsilon }}_{\pm \alpha }^{\dot{%
\alpha}}$), which is predicted by the solving of the constraint eq(\ref{con1}%
), see also (\ref{con2}), have been supplemented in order to cover the full
spectrum of 4D target space supersymmetry. \newline
(\textbf{2}) Borrow techniques from Penrose formalism to solve eq(\ref{con})
and then use the underlying free twistor like fields to build the fermionic
twistor like model with action $\mathcal{S}_{FT}=\mathcal{S}_{FT}\left[
\lambda \mathrm{,}\zeta \mathrm{,\mu ,\xi }\right] $ dual to the previous $%
\mathcal{S}_{F}$ one. Here $\lambda \mathrm{,}$ $\zeta \mathrm{,}$ $\mathrm{%
\mu }$ and $\mathrm{\xi }$ stand respectively for the twistor like world
fields $\lambda ^{\alpha }\left( \tau \right) \mathrm{,}$ $\zeta _{\pm
}^{\alpha }\left( \tau \right) $, $\mathrm{\mu }^{\dot{\alpha}}\left( \tau
\right) $\textrm{\ }and\textrm{\ }$\mathrm{\xi }_{\pm \dot{\alpha}}\left(
\tau \right) $ together with their canonical conjugate. These fields were
shown to be the projective coordinates of a supertwistor space whose
explicit structure depends on the signature of target space time geometry $%
\mathbb{R}^{\left( \mathrm{d,4-d}\right) }$. \newline
(\textbf{3}) Quantize canonically the twistor like free fields and solve the
underlying quantum gauge constraints to finish at the end of the analysis
with superfields describing $4D$ $\mathcal{N}=1$ target space supersymmetric
representations. Fermionic twistor like models with $4D$ extended
supersymmetry are also possible; they require implementation of more world
fermions.

Among the basic results of this study we would like to mention the two
following: \newline
(\textbf{i}) Penrose construction dealing with massless particles $%
x^{m}=x^{m}\left( \tau \right) $ moving in $4D$ space- time $\mathbb{R}%
^{\left( \mathrm{d,4-d}\right) }$ with energy momentum $p_{m}\sim $ \textrm{p%
}$_{\dot{\alpha}}^{\alpha }$ turns out to be apply to more general systems.
The point is that the usual Penrose incidence relation $\mathrm{\mu }^{\dot{%
\alpha}}=ix_{\beta }^{\dot{\alpha}}\lambda ^{\beta }$ eq(\ref{inc}) as well
as the representation,
\begin{equation}
\mathrm{p}_{\dot{\alpha}}^{\alpha }=\mathrm{\lambda }^{\alpha }\mathrm{%
\lambda }_{\dot{\alpha}},  \label{u1}
\end{equation}%
solving the condition \textrm{p}$^{2}=0$ have infinitely many analogues with
the desired features. For instance, eq(\ref{u1}) is nothing but the leading
term of a large class of world line fields belonging to the field family,%
\begin{equation}
\left( \mathrm{\Pi }_{\left( s_{1},s_{2}\right) }\right) _{\dot{\alpha}%
}^{\alpha }=\left( \mathrm{f}_{s_{1}}\right) ^{\alpha }\left( \mathrm{g}%
_{s_{2}}\right) _{\dot{\alpha}},\qquad s=s_{1}+s_{2}=0,\pm 1,\pm 2,...\text{
\ \ }.  \label{con11}
\end{equation}%
The $\mathrm{\Pi }_{\left( s_{1},s_{2}\right) }$ fields with $s_{1}s_{2}\neq
\left( 2n+1\right) $ capture all the required properties of $\mathrm{p}_{%
\dot{\alpha}}^{\alpha }$ needed for the building of twistor QFT. Like for $%
\mathrm{p}_{\dot{\alpha}}^{\alpha }$, the quadratic invariant $\mathrm{\Pi }%
_{\left( s_{1},s_{2}\right) }^{2}$ of these bispinors is given by
\begin{equation}
\mathrm{\Pi }_{\left( s_{1},s_{2}\right) }^{2}=\mathrm{f}_{s_{1}}^{2}.%
\mathrm{g}_{s_{2}}^{2},
\end{equation}%
and vanishes identically whenever at least one of the world line fields $%
\mathrm{f}_{s_{1}}=\mathrm{f}_{s_{1}}\left( \tau \right) $ or $\mathrm{g}%
_{s_{2}}=\mathrm{g}_{s_{2}}\left( \tau \right) $ is a boson. The next
leading terms in the above family coming after $\mathrm{p}_{\dot{\alpha}%
}^{\alpha }$, and which is associated with the level $s_{1}+s_{2}=0$,
corresponds obviously to taking,%
\begin{equation}
s_{1}+s_{2}=\pm 1.  \label{con2}
\end{equation}%
Thus we have the four following,%
\begin{eqnarray}
\left( \mathrm{\Pi }_{\left( \pm 1,0\right) }\right) _{\dot{\alpha}}^{\alpha
} &=&\mathrm{\zeta }_{\pm }^{\alpha }\text{ }\widetilde{\mathrm{\lambda }}_{%
\dot{\alpha}}\text{ \ \ \ }\sim \text{\ \ \ }\mathrm{\Upsilon }_{\pm \alpha
}^{\dot{\alpha}}  \notag \\
\left( \mathrm{\Pi }_{\left( 0,\pm 1\right) }\right) _{\dot{\alpha}}^{\alpha
} &=&\mathrm{\lambda }^{\alpha }\text{ }\widetilde{\mathrm{\zeta }}_{\pm
\dot{\alpha}}\text{ \ \ \ }\sim \text{\ \ }\widetilde{\mathrm{\Upsilon }}%
_{\pm \alpha }^{\dot{\alpha}}.  \label{con3}
\end{eqnarray}%
Because of the identities $\mathrm{\lambda }^{\alpha }\mathrm{\lambda }%
_{\alpha }=0$ and $\widetilde{\mathrm{\lambda }}_{\dot{\alpha}}\widetilde{%
\mathrm{\lambda }}^{\dot{\alpha}}=0$, these world line fields obey the Fermi
nilpotency property,%
\begin{equation}
\mathrm{Tr}\left( \mathrm{\Pi }_{\left( \pm 1,0\right) }^{2}\right)
=0,\qquad \mathrm{Tr}\left( \mathrm{\Pi }_{\left( 0,\pm 1\right)
}^{2}\right) =0,
\end{equation}%
capturing the known antisymmetry property of fermions wave function. In our
concern, these relations are thought of as the analogue of the condition
\textrm{p}$^{2}=0$ for massless particles in target space- time $\mathbb{R}%
^{\left( \mathrm{d,4-d}\right) }$, and so we can use Penrose method to deal
with them.\newline
(\textbf{ii}) Though the field action of the first level of the
construction, (see also (\ref{x4}-\ref{x8}),%
\begin{eqnarray}
\mathcal{S}_{F} &\sim &\int_{\mathbb{R}}d\tau \left[ \frac{\mathrm{1}}{2}%
\left( \mathrm{\Upsilon }_{-\alpha }^{\dot{\alpha}}\frac{\partial }{\partial
\tau }\mathrm{\Upsilon }_{+\beta }^{\dot{\beta}}-\mathrm{\Upsilon }_{+\beta
}^{\dot{\beta}}\frac{\partial }{\partial \tau }\mathrm{\Upsilon }_{-\alpha
}^{\dot{\alpha}}\right) \epsilon ^{\alpha \beta }\epsilon _{\dot{\alpha}\dot{%
\beta}}\right]  \notag \\
&&+\int_{\mathbb{R}}d\tau \left[ \frac{\mathrm{1}}{2}\left( \widetilde{%
\mathrm{\Upsilon }}_{-\alpha }^{\dot{\alpha}}\frac{\partial }{\partial \tau }%
\widetilde{\mathrm{\Upsilon }}_{+\beta }^{\dot{\beta}}-\widetilde{\mathrm{%
\Upsilon }}_{+\beta }^{\dot{\beta}}\frac{\partial }{\partial \tau }%
\widetilde{\mathrm{\Upsilon }}_{-\alpha }^{\dot{\alpha}}\right) \epsilon
^{\alpha \beta }\epsilon _{\dot{\alpha}\dot{\beta}}\right] ,  \label{con12}
\end{eqnarray}%
involves only world line fermions $\mathrm{\Upsilon }_{\pm \alpha }^{\dot{%
\alpha}}$ and $\widetilde{\mathrm{\Upsilon }}_{\pm \alpha }^{\dot{\alpha}}$
(no world line bosons $\mathrm{x}_{\alpha }^{\dot{\alpha}}$ nor its
conjugate momentum $\mathrm{p}_{\alpha }^{\dot{\alpha}}$ ), its twistor like
dual namely,%
\begin{eqnarray}
\mathcal{S}_{FT} &\sim &\int d\tau \left[ \frac{1}{2}\left( \mathrm{\nu }%
_{\alpha }\partial _{\tau }\lambda ^{\alpha }-\lambda ^{\alpha }\partial
_{\tau }\mathrm{\nu }_{\alpha }\right) -\frac{1}{2}\left( \mathrm{\nu }^{%
\dot{\alpha}}\partial _{\tau }\lambda _{\dot{\alpha}}-\lambda _{\dot{\alpha}%
}\partial _{\tau }\mathrm{\nu }^{\dot{\alpha}}\right) \right]  \notag \\
&&+\int d\tau \left[ \frac{1}{2}\left( \mathrm{\xi }_{-\alpha }\partial
_{\tau }\zeta _{+}^{\alpha }+\zeta _{+}^{\alpha }\partial _{\tau }\mathrm{%
\xi }_{-\alpha }\right) -\frac{1}{2}\left( \mathrm{\xi }_{-}^{\dot{\alpha}%
}\partial _{\tau }\zeta _{+\dot{\alpha}}+\zeta _{+\dot{\alpha}}\partial
_{\tau }\mathrm{\xi }_{-}^{\dot{\alpha}}\right) \right] .
\end{eqnarray}%
involves however bosonic ($\lambda ^{\alpha },$ $\mathrm{\nu }_{\alpha }$, $%
\lambda _{\dot{\alpha}}$, $\mathrm{\nu }^{\dot{\alpha}}$ ) and fermionic ($%
\zeta _{+}^{\alpha }$, $\mathrm{\xi }_{-\alpha }$, $\zeta _{+\dot{\alpha}}$,
$\mathrm{\xi }_{-}^{\dot{\alpha}}$) twistor like fields; thanks to the
splitting (\ref{con3}). This feature, which shows that eq(\ref{con12}) is
just a part of more general picture (see also last paragraph of this
discussion); should be viewed as a novelty in twistor field theory. The
reason is that starting with eq(\ref{con12}) and solve the underlying
constraint eqs, one discovers that the quantum spectrum of the dual
fermionic twistor like model has automatically $4D$ target space
supersymmetry.

\qquad In the end, we would like to note that the fermionic twistor
construction developed in this paper depends on the signature of 4D target
space $\mathbb{R}^{\left( d,4-d\right) }$ since corresponding spinors
exhibit different properties. In the present study we have given explicit
details for the case of particles moving in target space $\mathbb{R}^{\left(
2,2\right) }$; the extension to the others geometries is straightforward.
There, one also has the power of algebraic complex geometry. Note also that
our analysis applies as well to world sheet fields and heterotic $\left(
1,0\right) $\ and $\left( 0,1\right) $ QFT$_{2}$ models along the line
described in sections 2 and 3. Note moreover that, though enough to recover $%
4D$ target space time supersymmetry, our pure fermionic model is in fact not
completely independent from the standard Penrose bosonic one. Both of them
may be viewed as just two parts of a world line supersymmetric theory.
There, the fields $\mathrm{\Upsilon }_{\pm \alpha }^{\dot{\alpha}}$ can be
viewed as the fermionic partners of $x^{m}\sim x_{\alpha }^{\dot{\alpha}}$,
that is interchanged under world line supersymmetry, $\delta
_{susy}x_{\alpha }^{\dot{\alpha}}=\varepsilon _{\mp }\mathrm{\Upsilon }_{\pm
\alpha }^{\dot{\alpha}}$ and so on. This idea has been discussed shortly in
present paper (sections 2 and 3). It would be interesting to deeper this
issue for world line fields; but also for strings. Progress in this
direction will be reported elsewhere.

\begin{acknowledgement}
\qquad\ \ \newline
This research work is supported by the program Protars III D12/25, CNRST.
The authors thank ICTP (Senior Associate Programme and Organizers of Spring
School) for kind hospitality and where part of this work has been done.
\end{acknowledgement}

\end{document}